\documentclass[a4paper,fleqn,usenatbib]{mnras}

\usepackage{txfonts}

\usepackage[T1]{fontenc}
\usepackage{ae,aecompl}

\usepackage{graphicx}	
\usepackage{amssymb}	
\usepackage{longtable,lscape}

\newcommand{\AlII}{\hbox{{\rm Al}{\sc \,ii}}}
\newcommand{\AlIII}{\hbox{{\rm Al}{\sc \,iii}}}

\newcommand{\CaI}{\hbox{{\rm Ca}{\sc \,i}}}
\newcommand{\CaII}{\hbox{{\rm Ca}{\sc \,ii}}}

\newcommand{\CrII}{\hbox{{\rm Cr}{\sc \,ii}}}

\newcommand{\FeII}{\hbox{{\rm Fe}{\sc \,ii}}}

\newcommand{\HI}{\hbox{{\rm H}{\sc \,i}}}

\newcommand{\MgI}{\hbox{{\rm Mg}{\sc \,i}}}
\newcommand{\MgII}{\hbox{{\rm Mg}{\sc \,ii}}}

\newcommand{\ZnII}{\hbox{{\rm Zn}{\sc \,ii}}}

\newcommand{\kms}{\hbox{km~s$^{-1}$}}

\newcommand{\To}{\hbox{$T_{0}$}}
\newcommand{\Tn}{\hbox{$T_{90}$}}

\title[GRB\,110715A]{GRB\,110715A: The peculiar multiwavelength evolution of the first afterglow detected by ALMA
\thanks{This publication is based on data acquired with the Atacama Pathfinder Experiment (APEX) under program 087.F-9301(A). This paper makes use of the following ALMA data: ADS/JAO.ALMA\#2011.0.00001.E. This publication is based on data acquired with the VLT/X-shooter under program 087.A-0055(C), as well as with VLT/FORS2 under program 091.A-0703(A).}}

\author[R.~S\'{a}nchez-Ram\'{\i}rez et al.]{
R.~S\'{a}nchez-Ram\'{\i}rez,$^{1,2,3}$\thanks{E-mail: ruben@iaa.es}
P.~J.~Hancock,$^{4,5}$
G.~J\'{o}hannesson,$^{6}$
Tara~Murphy,$^{4}$\newauthor
A.~de~Ugarte~Postigo,$^{1,7}$
J.~Gorosabel,$^{1,2,3}$\thanks{Deceased}
D.~A.~Kann,$^{8,9}$
T.~Kr\"{u}hler,$^{9,10}$\newauthor
S.~R.~Oates,$^{1,11}$
J.~Japelj,$^{12}$
C.~C.~Th\"{o}ne,$^{1}$
A.~Lundgren,
D.~A.~Perley,$^{7}$\newauthor
D.~Malesani,$^{7}$
I.~de~Gregorio~Monsalvo,$^{10,13}$
A.~J.~Castro-Tirado,$^{1}$
V.~D'Elia,$^{14,15}$\newauthor
J.~P.~U.~Fynbo,$^{7}$
D.~Garcia-Appadoo,$^{10,13}$
P.~Goldoni,$^{16}$
J.~Greiner,$^{9,17}$
Y.-D. Hu,$^{1}$\newauthor
M.~Jel\'{i}nek,$^{1,18}$
S.~Jeong,$^{1,19}$
A.~Kamble,$^{20}$
S.~Klose,$^{8}$
N.~P.~M.~Kuin,$^{11}$
A.~Llorente,$^{21}$\newauthor
S.~Mart\'{i}n,$^{10,13}$
A.~Nicuesa~Guelbenzu,$^{8}$
A.~Rossi,$^{22}$
P.~Schady,$^{9}$
M.~Sparre,$^{7}$\newauthor
V.~Sudilovsky,$^{23}$
J.~C.~Tello,$^{1}$
A.~Updike,$^{24}$
K.~Wiersema$^{25}$
and B.-B. Zhang$^{1,26}$
\\
\\
Affiliations are listed at the end of the paper
}

\date{Accepted XXX. Received YYY; in original form ZZZ}

\pubyear{2016}

\begin{document}
\label{firstpage}
\pagerange{\pageref{firstpage}--\pageref{lastpage}}
\maketitle

\begin{abstract}
We present the extensive follow-up campaign on the afterglow of GRB\,110715A at 17 different wavelengths, from X-ray to radio bands, starting 81 seconds after the burst and extending up to 74 days later. We performed for the first time a GRB afterglow observation with the ALMA observatory. We find that the afterglow of GRB\,110715A is very bright at optical and radio wavelengths. We use optical and near infrared spectroscopy to provide further information about the progenitor's environment and its host galaxy. The spectrum shows weak absorption features at a redshift $z$ = 0.8225, which reveal a host galaxy environment with low ionization, column density and dynamical activity. Late deep imaging shows a very faint galaxy, consistent with the spectroscopic results. The broadband afterglow emission is modelled with synchrotron radiation using a numerical algorithm and we determine the best fit parameters using Bayesian inference in order to constrain the physical parameters of the jet and the medium in which the relativistic shock propagates. We fitted our data with a variety of models, including different density profiles and energy injections. Although the general behaviour can be roughly described by these models, none of them are able to fully explain all data points simultaneously. GRB\,110715A shows the complexity of reproducing extensive multi-wavelength broadband afterglow observations, and the need of good sampling in wavelength and time and more complex models to accurately constrain the physics of GRB afterglows.
\end{abstract}




\section{Introduction}

Gamma-ray bursts \citep[GRBs,][]{1973ApJ...182L..85K} are the most violent explosions in the Universe. They are characterized by a short flash of gamma-ray photons followed by a long lasting afterglow that can be observed at all wavelengths. They can be classified into two types based on the duration (and the hardness) of their $\gamma$-emission: short and long GRBs \citep[\Tn\ $<$ 2 s (hard spectrum) and \Tn\ $>$ 2 s (soft spectrum) respectively;][]{1993ApJ...413L.101K}. Currently, the most favored model to explain the origin of GRBs is a highly magnetized relativistic jet, but more prompt polarimetric observations are needed in order to confirm this \citep{2013Natur.504..119M,2015ApJ...813....1K}. The prompt emission likely originates from either internal shocks in the photosphere of the jet or magnetic dissipation in a magnetically dominated jet \citep[see][and references therein]{2011ApJ...726...90Z,2012ApJ...751...90Z}. The afterglow emission, however, is thought to originate from external shocks caused by the jet's interaction with the interstellar medium (ISM). Multiwavelength emission is expected to be produced by a forward shock moving into the ISM and a reverse shock moving into the expanding jet \citep{1993ApJ...405..278M, 1998ApJ...497L..17S,1999ApJ...519L..17S}. The reverse shock is supposed to be short lived, with most of the afterglow emission being generated by the forward shock.

The electrons accelerated at the shock fronts emit synchrotron radiation as they interact with the magnetic field behind the shock fronts. By modeling this emission we can determine the  physical parameters of the GRB ejecta and the structure of the ISM near the progenitor along the line of sight to Earth. The most popular way to extract the parameters is by using an analytical model for the expected shape of the afterglow light curves and spectrum \citep{1997ApJ...487L...1R,1998ApJ...497L..17S,1999ApJ...523..177W}. The emission is split into regions in time and wavelength, where the resulting light curve and spectrum can be approximated by power laws. The slopes of these segments along with the location of the spectral breaks are then used to determine the physical parameters. An alternative method is to model the emission using a numerical code that takes as input the physical parameters of interest \citep{2002ApJ...571..779P,  2006ApJ...647.1238J}. In  addition to requiring fewer approximations, the numerical models allow us to study a more complex structure for the ISM and the GRB ejecta and is therefore our method of choice for this study.

To properly determine the physical properties of the GRB ejecta a wide range of accurate multi-wavelength observations are needed with as good time coverage as possible. The millimetre/sub-millimetre range is of crucial importance in constraining the afterglow models as it is where the flux density of the emission peaks during the first few days after the GRB onset. In  this range the capabilities of the new ALMA observatory bring an important leap forward, thanks to its great improvement in resolution and sensitivity in comparison with previous observatories \citep{2012A&A...538A..44D}.

It is widely accepted that long GRBs are created by the explosive death of massive stars \citep{2003Natur.423..847H, 2006ARA&A..44..507W}, probably rapidly rotating Wolf-Rayets \citep[for a review see, e.g.,][]{2007ARA&A..45..177C}. However, it remains unclear what the specific mechanism in the core collapse process is that triggers the formation of a jet.  Given the short life periods of such massive stars and their luminosity, GRB afterglows can be used as powerful tracers of star-forming galaxies over a wide range of redshifts \citep[e.g.,][]{2009ApJ...705L.104K, 2012ApJ...744...95R, 2015A&A...581A.125K, 2015ApJ...808...73S, 2015A&A...581A.102V}. To date spectroscopically confirmed GRB redshifts range from $z$ = 0.0085 \citep[GRB\,980425A][]{1998IAUC.6896....3W,1998Natur.395..670G} to $z$ = 8.2 \citep[GRB\,090423A][]{2009Natur.461.1254T, 2009Natur.461.1258S}, and photometric redshifts have been proposed up to $z$ = 9.4 \citep[GRB\,090429B][]{2011ApJ...736....7C}. 

Optical/NIR spectroscopy of GRB afterglows can also be used to study the intervening matter present along the line of sight at different distance scales, ranging from regions around the progenitor to distant intervening galaxies. Its strength resides in the extremely bright afterglow, making it possible to measure atomic/molecular transitions in the host galaxy and intervening systems \citep[e.g.][]{2014A&A...564A..38D}, even when the probed galaxies are not detected by deep direct imaging. It made possible to accurately probe absorption metalliticity out to $z$ $\geq$ 5 \citep[e.g.,][]{2014ApJ...785..150S, 2015A&A...580A.139H}.

In this paper we present observations of the afterglow and host galaxy of GRB\,110715A, aiming at understanding the physical processes involved in the explosion and the environment in which it occurred. Our data includes radio, submillimeter, near-infrared, optical, ultraviolet, and X-ray observations. In Section 2 we introduce the observations available for this event, in Section 3 we explain the results of these observations and discuss the implications, and in Section 4 we present our conclusions.

\section{Observations}

\begin{figure*}
\includegraphics[width=\textwidth]{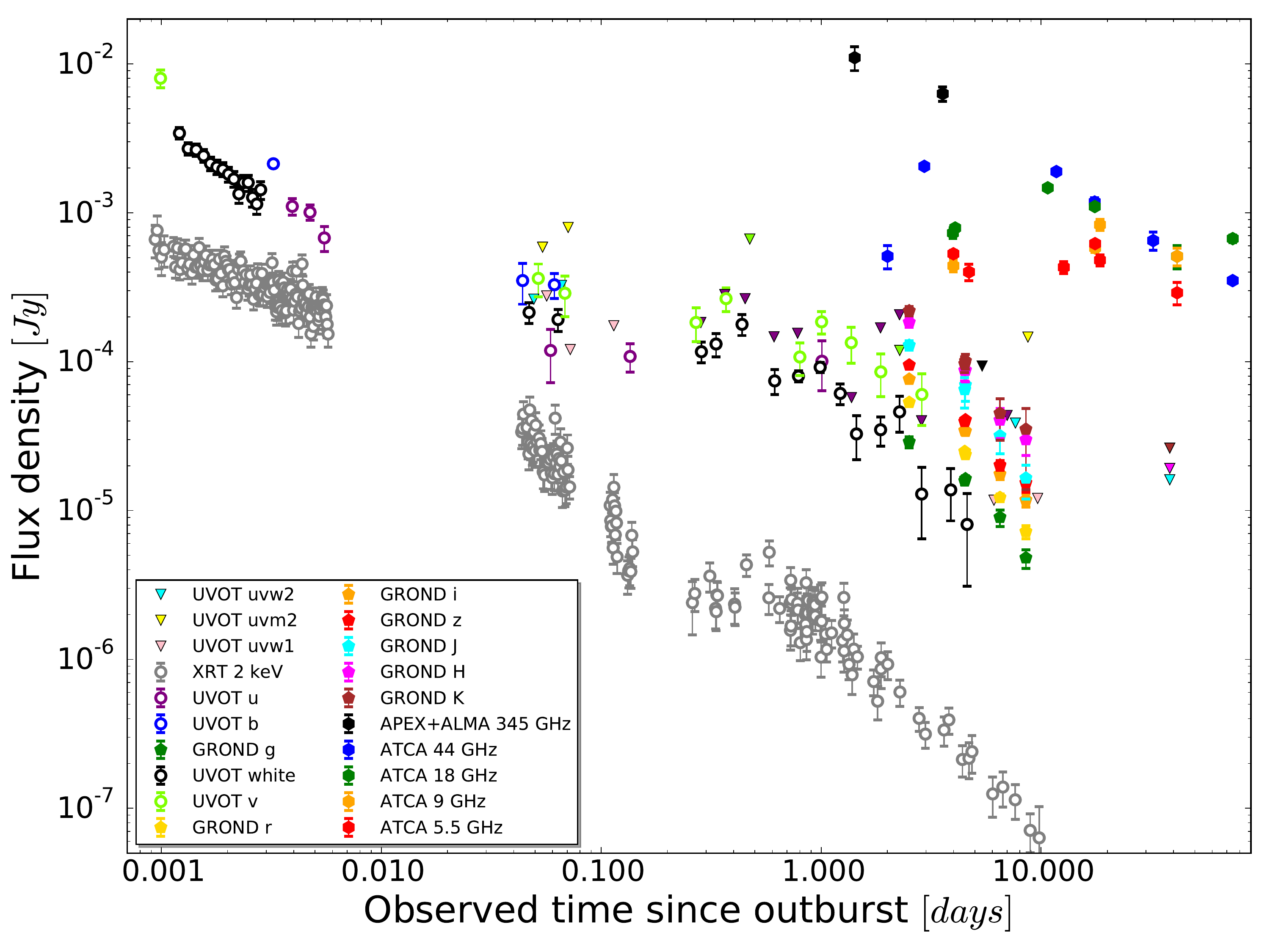}
\caption{Afterglow light curve (Galactic extinction corrected) of the 17 bands observed. Upper limits are denoted by down-pointing triangles.}
\label{fig:obs}
\end{figure*}

\begin{table}
\caption{Broad band multiwavelength observations of GRB\,110715A. The full table is available in the online version.}
\centering
\begin{tabular}{cccc}
\hline\hline
$T$ - $T_{0}$ & Flux & AB & Band \\
days & Jy & mag & \\
\hline
0.00094 & (1.32 $\pm$ 0.33) $\times$ 10$^{-04}$ & 18.60$^{+0.24}_{-0.31}$ & XRT 2 keV \\ 
0.00096 & (1.52 $\pm$ 0.37) $\times$ 10$^{-04}$ & 18.44$^{+0.24}_{-0.31}$ & XRT 2 keV \\ 
0.00098 & (1.12 $\pm$ 0.27) $\times$ 10$^{-04}$ & 18.78$^{+0.24}_{-0.31}$ & XRT 2 keV \\ 
0.00100 & (1.01 $\pm$ 0.25) $\times$ 10$^{-04}$ & 18.89$^{+0.24}_{-0.31}$ & XRT 2 keV \\ 
0.00103 & (1.13 $\pm$ 0.26) $\times$ 10$^{-04}$ & 18.77$^{+0.23}_{-0.29}$ & XRT 2 keV \\ 
0.00114 & (1.19 $\pm$ 0.17) $\times$ 10$^{-04}$ & 18.71$^{+0.15}_{-0.17}$ & XRT 2 keV \\ 
0.00116 & (0.86 $\pm$ 0.13) $\times$ 10$^{-04}$ & 19.06$^{+0.16}_{-0.18}$ & XRT 2 keV \\ 
0.00119 & (1.16 $\pm$ 0.17) $\times$ 10$^{-04}$ & 18.74$^{+0.15}_{-0.18}$ & XRT 2 keV \\ 
... & ... & ... & ... \\ 
\hline
\end{tabular}
\label{tab:obs}
\end{table}

\subsection{Gamma-ray emission}

The \textit{Swift} \citep{2004ApJ...611.1005G} Burst Alert Telescope \citep[BAT,][]{2005SSRv..120..143B} triggered and located GRB\,110715A on 15 July 2011 at \To = 13:13:50 UT \citep{2011GCN..12158...1S}. The gamma-ray light curve shows a double-peaked structure with a duration of \Tn = 13.0 $\pm$ 4.0 s (90\% confidence level) in the observer frame. Therefore we classify GRB\,110715A as a long burst.

Analysis of the time-integrated spectrum from \textit{Konus-Wind} gave the best fit as a Band function with the following parameters: $\alpha_1$ = -1.23$^{+0.09}_{-0.08}$, $\alpha_2$ = -2.7$^{+0.2}_{-0.5}$, $E_p$ = 120$^{+12}_{-11}$ \citep{2011GCN..12166...1G}. GRB\,110715A was also detected by \textit{INTEGRAL}/SPI-ACS,  and \textit{Suzaku}/WAM  \citep[see more details in][]{2011GCN..12166...1G,2011GCN..12158...1S}.

\subsection{X-ray afterglow observations}

The X-Ray Telescope \citep[XRT;][]{2005SSRv..120..165B} onboard {\it Swift} began observing the field 90.9 seconds after the BAT trigger, localizing the X-ray afterglow at RA(J2000) = 15$h$ 50$m$ 44.00$s$, Dec.(J2000) = -46$^{\circ}$  14$\arcmin$ 07\farcs5 with an uncertainty of 1\farcs4 \citep[90\% confidence level;][]{2011GCN..12161...1E}. 

The unabsorbed X-ray afterglow light curve used in this paper has been extracted from the Burst Analyzer \footnote{\url{http://www.swift.ac.uk/burst_analyser}} \citep{2007A&A...469..379E,2009MNRAS.397.1177E,2010A&A...519A.102E}, which uses the spectral slope to derive the flux densities at an energy of 2 keV and assumes a Hydrogen column density N(H) = 1.6$^{+0.5}_{-0.4}$ $\times$ 10$^{22}$ cm$^{-2}$. These observations are shown in Figure \ref{fig:obs} and tabulated in Table \ref{tab:obs}.

\subsection{UV/Optical/NIR afterglow observations}

\begin{table}
\caption{Effective wavelengths and extinction coefficients.}
\begin{tabular}{lcc}
\hline\hline
Band & $\lambda_{\rm eff}$ [$\mu$m] & A$_{\lambda}$ $^{(a)}$ \\
\hline
UVOT uvw2 & 0.193 & 4.099 \\
UVOT uvm2 & 0.225 & 4.582 \\
UVOT uvw1 & 0.260 & 3.623 \\
UVOT u & 0.351 & 2.587 \\
UVOT b & 0.441 & 2.021 \\
GROND g' & 0.459 & 2.018 \\
UVOT \textit{white} & 0.483 & 2.566 \\
UVOT v & 0.545 & 1.628 \\
GROND r' & 0.622 & 1.393 \\
GROND i' & 0.764 & 1.042 \\
FORS2 Ic & 0.786 & 0.949 \\
GROND z' & 0.899 & 0.775 \\
GROND J & 1.239 & 0.455 \\
GROND H & 1.646 & 0.291 \\
GROND K & 2.170 & 0.187 \\
\hline
\end{tabular}
\\[0.2cm]
\begin{scriptsize}
(a) $E(B-V)$~=~0.52 mag \citep{2011ApJ...737..103S}
\end{scriptsize}
\label{tab:extinction}
\end{table}

GRB\,110715A was followed up in UV/Optical/NIR wavelengths with \textit{Swift} (+UVOT) and the 2.2m MPG telescope (+GROND). Light curves are shown in Figure \ref{fig:obs} and tabulated in Table \ref{tab:obs} as well.

This burst had a very bright optical counterpart in spite of the high Galactic extinction caused by its location close to the Galactic plane \citep{2011GCN..12161...1E}. The GRB afterglow study was affected by the Galactic reddening, initially estimated to be $E(B-V)$ = 0.59 mag according to the dust maps of \citet{1998ApJ...500..525S}, and later $E(B-V)$ = 0.52 mag following \citet{2011ApJ...737..103S}. We adopted the latest value. Computed effective wavelengths and extinction for each band are presented in Table \ref{tab:extinction}.

\subsubsection{UVOT imaging}

The \textit{Swift} Ultra-Violet/Optical Telescope \citep[UVOT;][]{2005SSRv..120...95R} began settled observations of the field of GRB\,110715A 100$s$ after the trigger \citep{2011AIPC.1358..373B}. The afterglow was detected in the $white$, $u$, $b$ and $v$ filters at RA(J2000) = 15h 50m 44.09s, Dec.(J2000) = -46$^{\circ}$  14$\arcmin$ 06\farcs5, with a $2\sigma$ uncertainty of about 0\farcs62. For this analysis, we have reduced both image and event mode data grouped with binning $\Delta t / t \sim 0.2$. Before the count rates were extracted from the event lists, the astrometry was refined following the  methodology in \citet{2009MNRAS.395..490O}. The photometry was then extracted from the event lists and image files based on the FTOOLs \textit{uvotevtlc} and \textit{uvotmaghist}, respectively, using a source aperture centered on the optical position and a background region located in a source-free zone. We used a 3" source aperture to avoid contamination from neighbouring stars and  applied aperture corrections to the photometry in order to be compatible with the UVOT calibration \citep{2011AIPC.1358..373B}. The analysis pipeline used software HEADAS 6.10 and UVOT calibration 20111031.

\subsubsection{GROND imaging}

We obtained follow-up observations of the optical/NIR afterglow of GRB\,110715A with the seven-channel imager GROND \citep[Gamma-ray burst optical/near-infrared detector;][]{2008PASP..120..405G} mounted on the 2.2m MPG@ESO telescope stationed in La Silla, Chile. The first observations were obtained 2.5 days after the trigger, after losing the first two nights due to weather. This first epoch suffers from very bad  seeing, 1\farcs5 -- 1\farcs9 depending on the band, but the optical/NIR afterglow was clearly detected \citep{2011GCN..12169...1U}.  Deeper follow-up  under  better conditions  in
three further  epochs reveals a faint nearby source which exhibits a stellar PSF. The presence of this source was carefully accounted for during the data analysis. The GROND optical and NIR image reduction and photometry were performed by calling on standard IRAF tasks \citep{1993ASPC...52..173T} using the custom GROND pipeline \citep{2008AIPC.1000..227Y}, similar to the procedure described in \cite{2008ApJ...685..376K}. Hereby, we used SExtractor \citep{1996A&AS..117..393B} for background modeling, and bright sources were masked out, which yields improved results in the case of this crowded field. A late epoch was obtained 38 days after the GRB which was supposed to be used for image subtraction purposes, but a positioning
error led to the afterglow position being covered only in the NIR frames.

Afterglow magnitudes in the optical were measured against comparison stars calibrated to the SDSS catalogue \citep{2011ApJS..193...29A}, obtained from observing an SDSS field at
similar airmass immediately after the fourth epoch, in photometric conditions. NIR magnitudes were measured against on-chip comparison stars taken from the 2MASS catalogue \citep{2006AJ....131.1163S}. The results of the photometry are displayed in Table \ref{tab:obs}.

\subsection{Submm afterglow observations}

\begin{figure}
\centering
\includegraphics[width=\columnwidth]{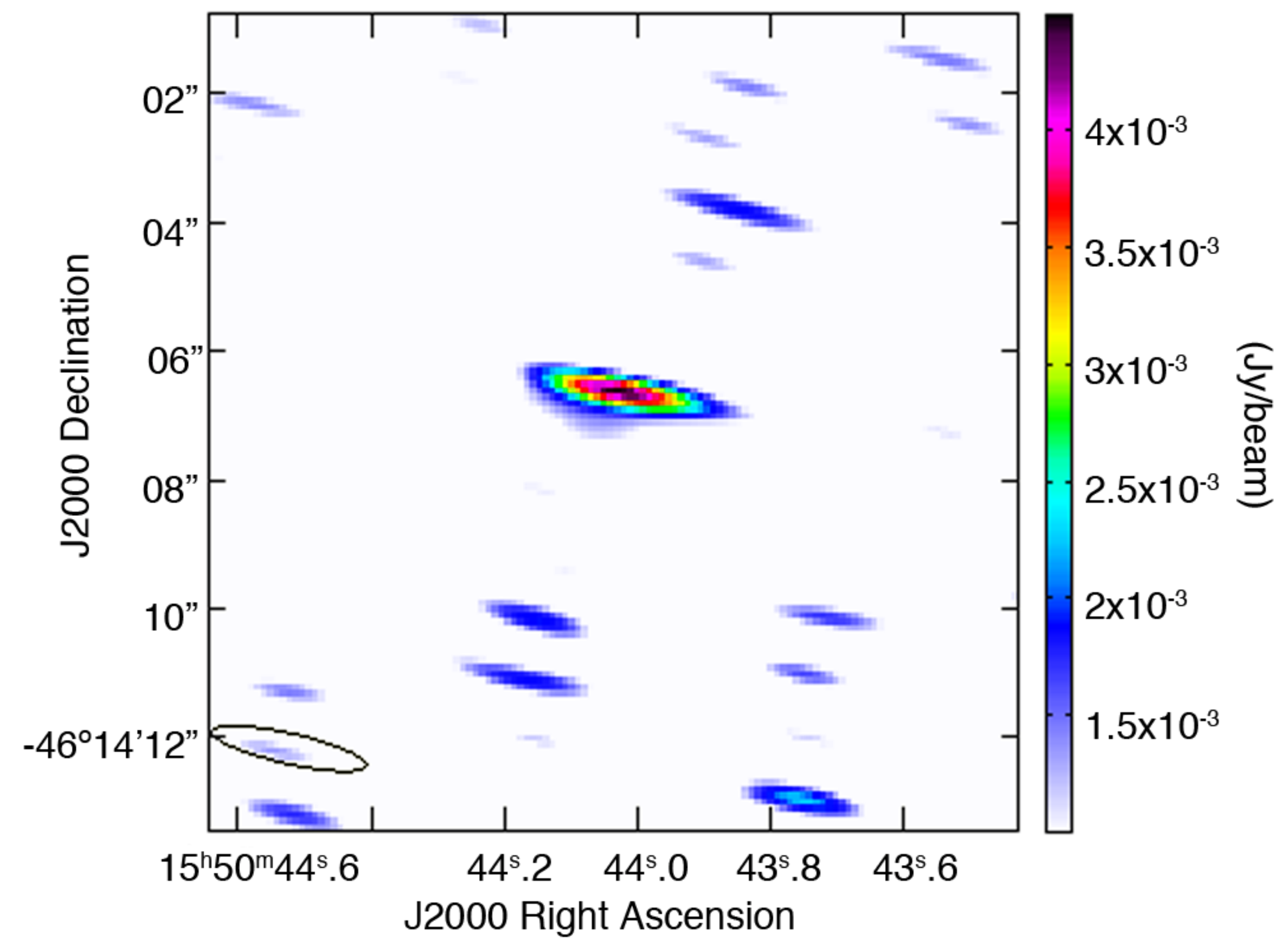}
\caption{ALMA image at 345 GHz. The beam size (0\farcs3 $\times$ 0\farcs1, P.A.=76 degrees) is showed in the lower left corner.}
\label{fig:alma}
\end{figure}

The Atacama pathfinder experiment telescope (APEX) observations began on July 16, 1.42 days after the burst and were performed in the 345 GHz band using the photometric mode of the Large Apex BOlometer CAmera \citep[LABOCA;][]{2009A&A...497..945S} under good weather conditions. Data reduction was done using the BoA (1), CRUSH and miniCRUSH \citep{2008SPIE.7020E..1SK} software packages. Using these observations we discovered a bright submm counterpart at 10.4 $\pm$ 2.4 mJy \citep{2011GCN..12168...1D}.

As a test of the target of opportunity programme, GRB\,110715A was also observed with the Atacama Large Millimeter Array (ALMA), yielding a detection with a flux density of 4.9 $\pm$  0.6 mJy at 345~GHz \citep{2012A&A...538A..44D}. The ALMA observations began on July 19 at 02:50 UT (3.57 days after the burst), and they were carried out making use of only 7 antennas during 25 mins on source. We present the data in Figure \ref{fig:alma}.

In spite of being obtained during a test observation, with almost an order of magnitude fewer antennas than are available with the full observatory, this was the deepest observation carried out to date at 345~GHz of a GRB afterglow \citep{2012A&A...538A..44D}. The ALMA observation also provides the most accurate coordinates available for this GRB. The centroid of the afterglow is located at RA(J2000) = 15$h$  50$m$  44.05$s$ and Dec.(J2000) = -46$^{\circ}$  14$\arcmin$ 06\farcs5 with a synthesized beam size of 0\farcs3 $\times$ 0\farcs1 at a position angle of 76 degrees, which provides an astrometric accuracy $\lesssim$ 0\farcs02. The differences between the optical and radio reference frames only limit this precision to be $\lesssim$ 0\farcs05 \citep[e.g.,][]{1986HiA.....7..103J}.

\subsection{Radio afterglow observations}

Following the detection of an afterglow at submm wavelengths with APEX \citep{2011GCN..12168...1D}, radio observations were obtained with the Australia Telescope Compact Array \citep[][ATCA]{2011MNRAS.416..832W} two and three days after the trigger. These observations resulted in further detections of the afterglow at $44$\,GHz \citep{2011GCN..12171...1H}. This GRB was monitored at 44, 18, 9, and 5 GHz for up to 75 days post-burst, where the flux remained at a sub-mJy level. The lower frequency observations were complicated by the presence of a second source within the field of view (MGPS\,J155058-461105). The data were reduced using standard procedures in MIRIAD \citep{1995ASPC...77..433S}. An additional late-time visit was performed on 12 Aug 2013 at 5.5\,GHz and 9\,GHz, to understand the possible contribution of the host galaxy, which was found to be negligible at both bands. The flux evolution of the afterglow at the four ATCA frequencies is also shown in Figure \ref{fig:obs} and tabulated in Table \ref{tab:obs}, together with the rest of the observing bands.

\subsection{Optical/nIR afterglow spectra}

\begin{table}
\caption{X-shooter observations log.}
\begin{tabular}{lllll}
\hline\hline
Mean T-\To & Arm & Exp. time  & Slit width & Resolution $(a)$ \\
(hr)		&	&	(s)	&	(\arcsec)	& \\
\hline
12.60 & UVB & 618.02 & 1.0 & 4350 \\
12.60 & VIS & 612.04 & 0.9 & 7450 \\
12.60 & NIR & 600.00 & 0.9 $(b)$ & 5300 \\
\hline
\end{tabular}
\\[0.2cm]
\begin{scriptsize}
(a) Nominal values.
(b) K-band blocker was not used.
\end{scriptsize}
\label{tab:xshlog}
\end{table}

\begin{figure*}
\centering
\includegraphics[width=0.85\textwidth]{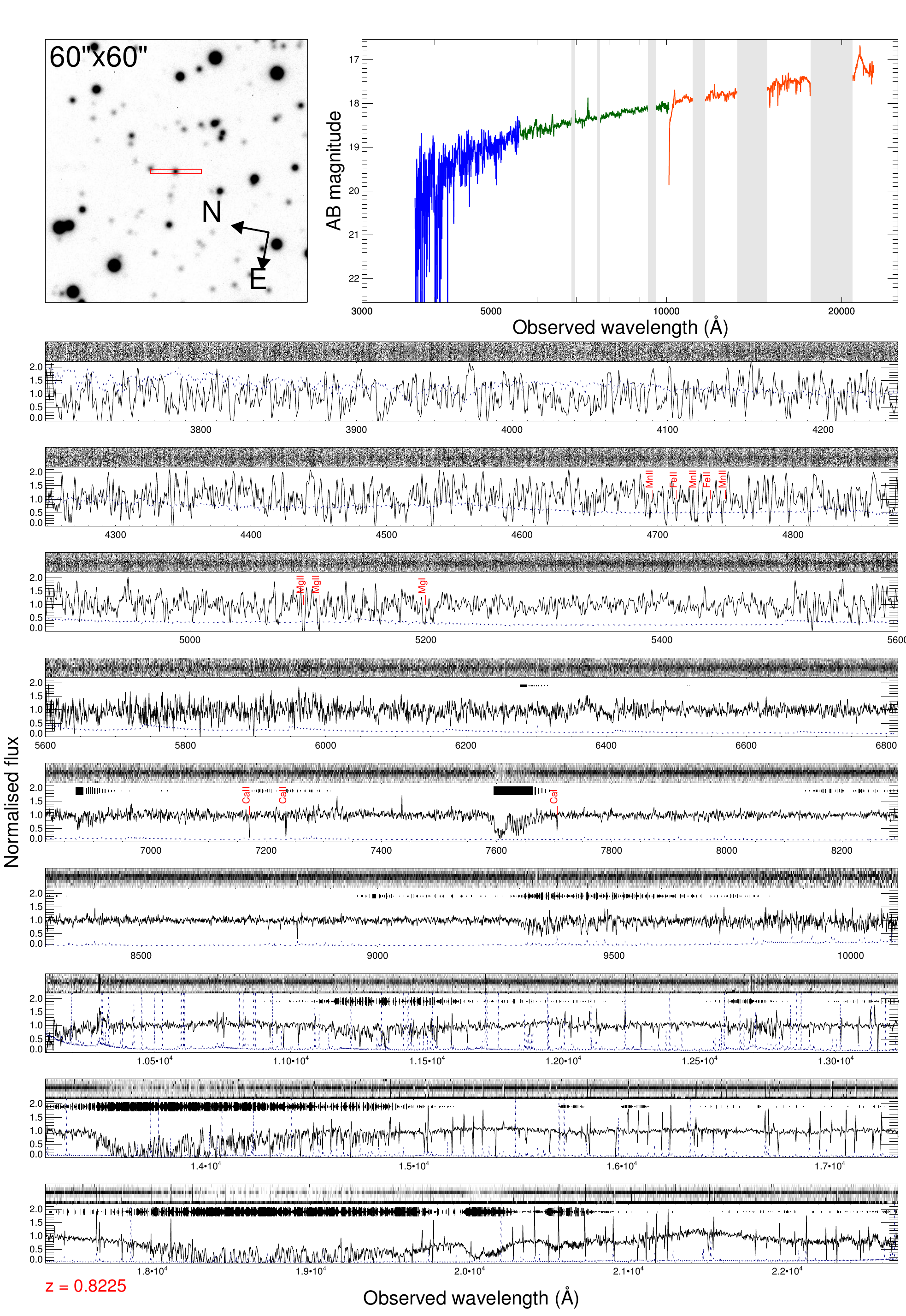}
\caption{X-shooter spectra. Upper panels are the finding chart (left) and an overview of the complete flux calibrated spectra, corrected for Galactic extinction (right). In the bottom plot, we show the normalised spectra, with 3 panels per arm, starting at top with UVB and followed by VIS and NIR. Each panel includes the 2D image and 1D signal and error spectrum. Telluric absorptions are indicated by black bands above the 1D spectrum, their thickness is a measure of the absorption strength.}
\label{fig:spec}
\end{figure*}

VLT/X-shooter \citep{2011A&A...536A.105V}, an optical/nIR intermediate resolution spectrograph mounted at the Very Large Telescope (VLT) Unit Telescope (UT) 2 in Paranal Observatory (Chile), was used to observe the GRB afterglow starting 12.7 hrs after the \textit{Swift} trigger. The seeing was 0\farcs9, but observations had to be interrupted due to wind constraints \citep{2011GCN..12164...1P}. The observing log is shown in Table \ref{tab:xshlog}. We processed the spectra using version 2.0.0 of the X-shooter data reduction pipeline \citep{2006SPIE.6269E..2KG, 2010SPIE.7737E..28M}. As the observations were stopped after one exposure, the standard nodding reduction could not be performed. We thus reduced the single frames of each arm with the following steps: We performed bias subtraction, cosmic ray detection and subtraction \citep{2001PASP..113.1420V}, and flat field division on the raw frames. From these processed frames the sky emission was subtracted using the \citet{2003PASP..115..688K} method and 1D spectra were extracted directly order by order from the sky-subtracted and flat-field divided frame using optimal extraction \citep{1986PASP...98..609H}. The resulting spectra were merged weighting them by the errors and the final merged spectra were then averaged in IDL. The spectra were flux calibrated using observations of the standard star LTT7987 taken the same night. The complete X-shooter spectrum is shown in Fig.~\ref{fig:spec}.

\subsection{Host galaxy imaging}

606 days after the burst, the field of GRB\,110715A was revisited using GROND searching for a possible host galaxy contribution. However, the data did not reveal any underlying source. We therefore derive only detection limits.

A deeper exposure was obtained on August 2013 with FORS2 at ESO's VLT 751 days after the burst. The observation consisted of $10\times240$ s in $I_C$-band, with a seeing of 0\farcs55, and data were reduced in a similar fashion as the GROND imaging. An object is detected close to the afterglow position at a magnitude of 26.40 $\pm$ 0.36 mag.

\begin{figure}
\centering
\includegraphics[width=\columnwidth]{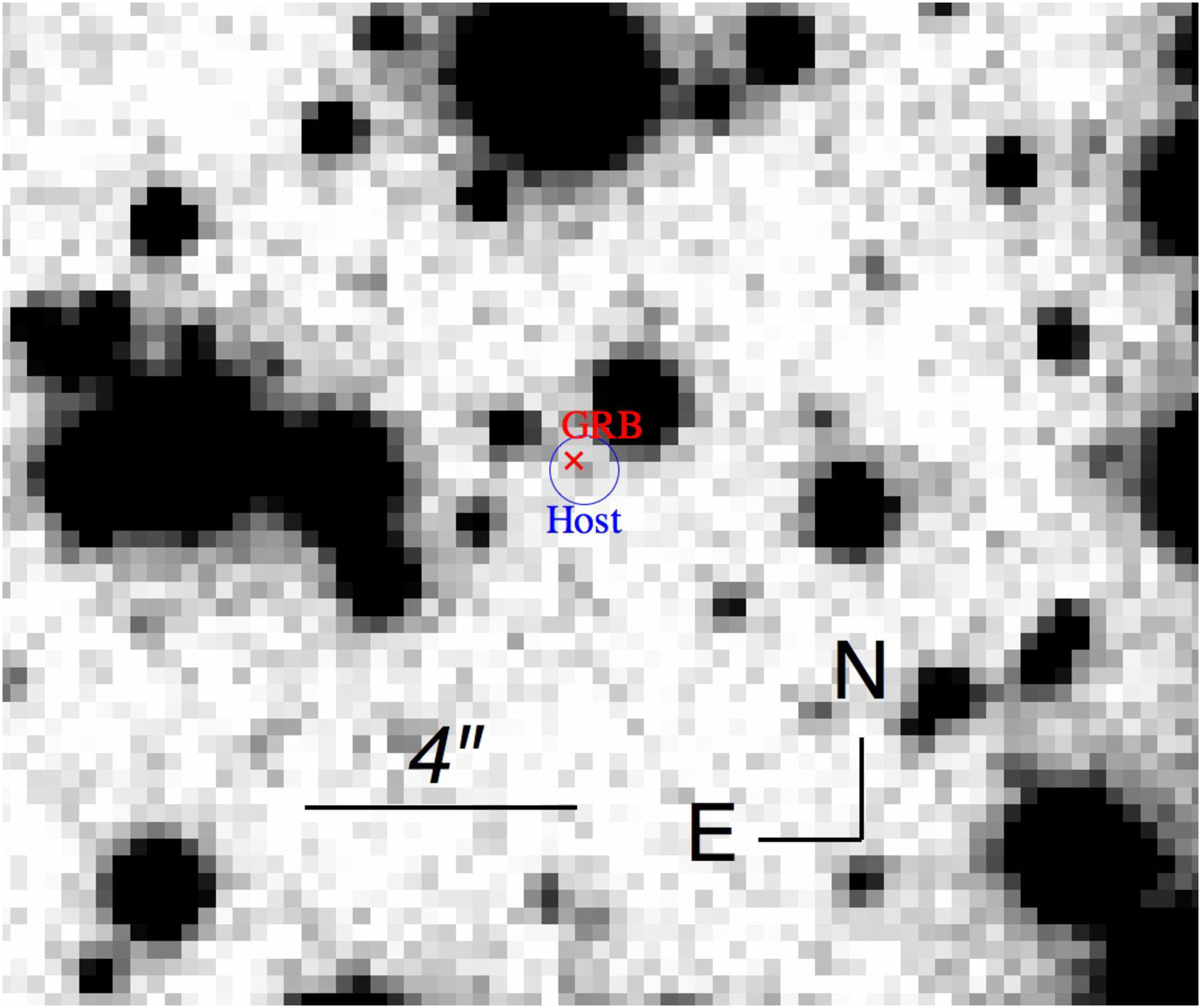}
\caption{Observation of the host galaxy in the $I_C$-band from VLT/FORS2.}
\label{fig:host}
\end{figure}

\section{Results and discussion}

\subsection{The afterglow of GRB\,110715A in a global context}

\begin{figure*}
\centering
\includegraphics[width=\columnwidth]{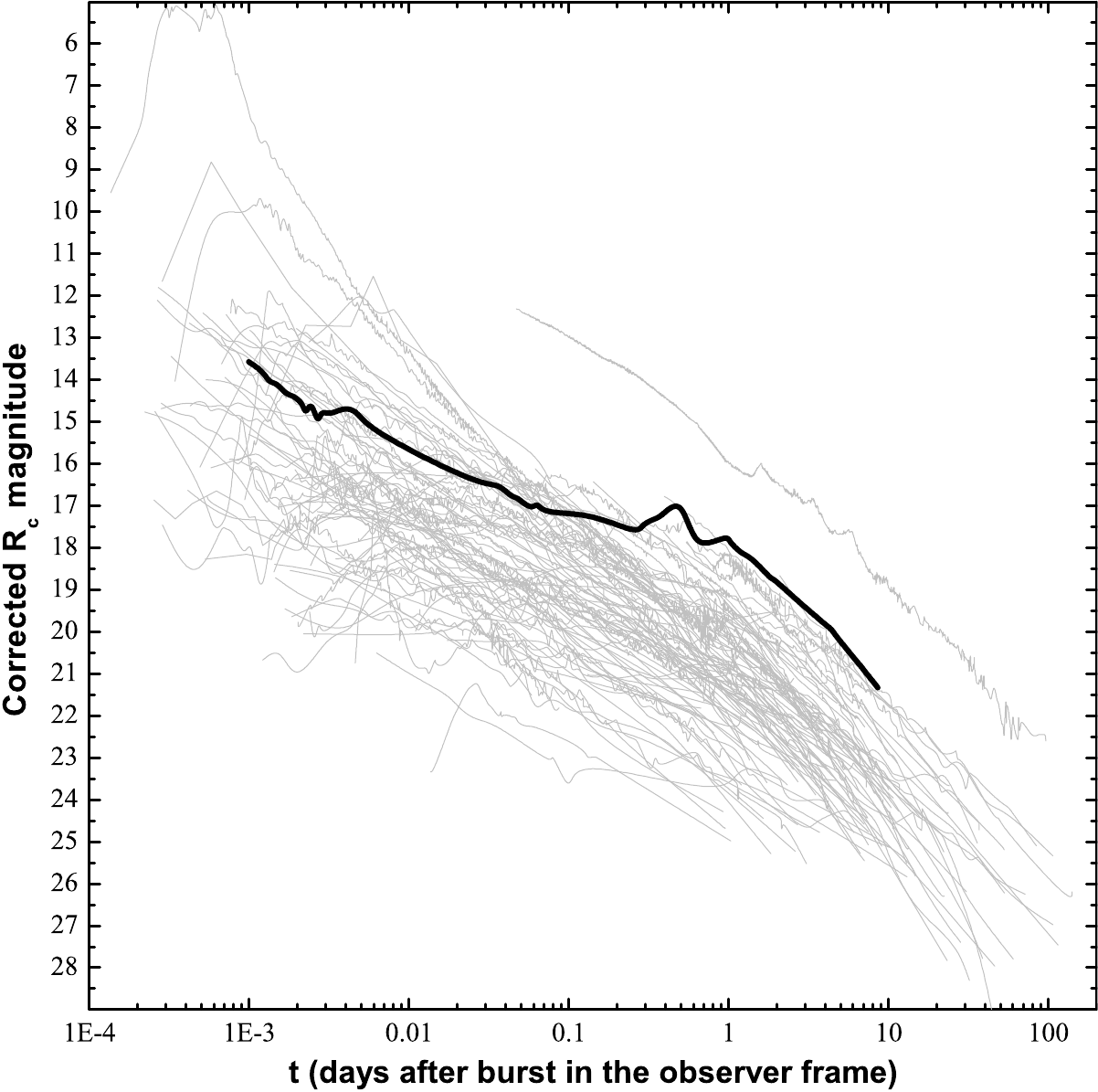}
\includegraphics[width=\columnwidth]{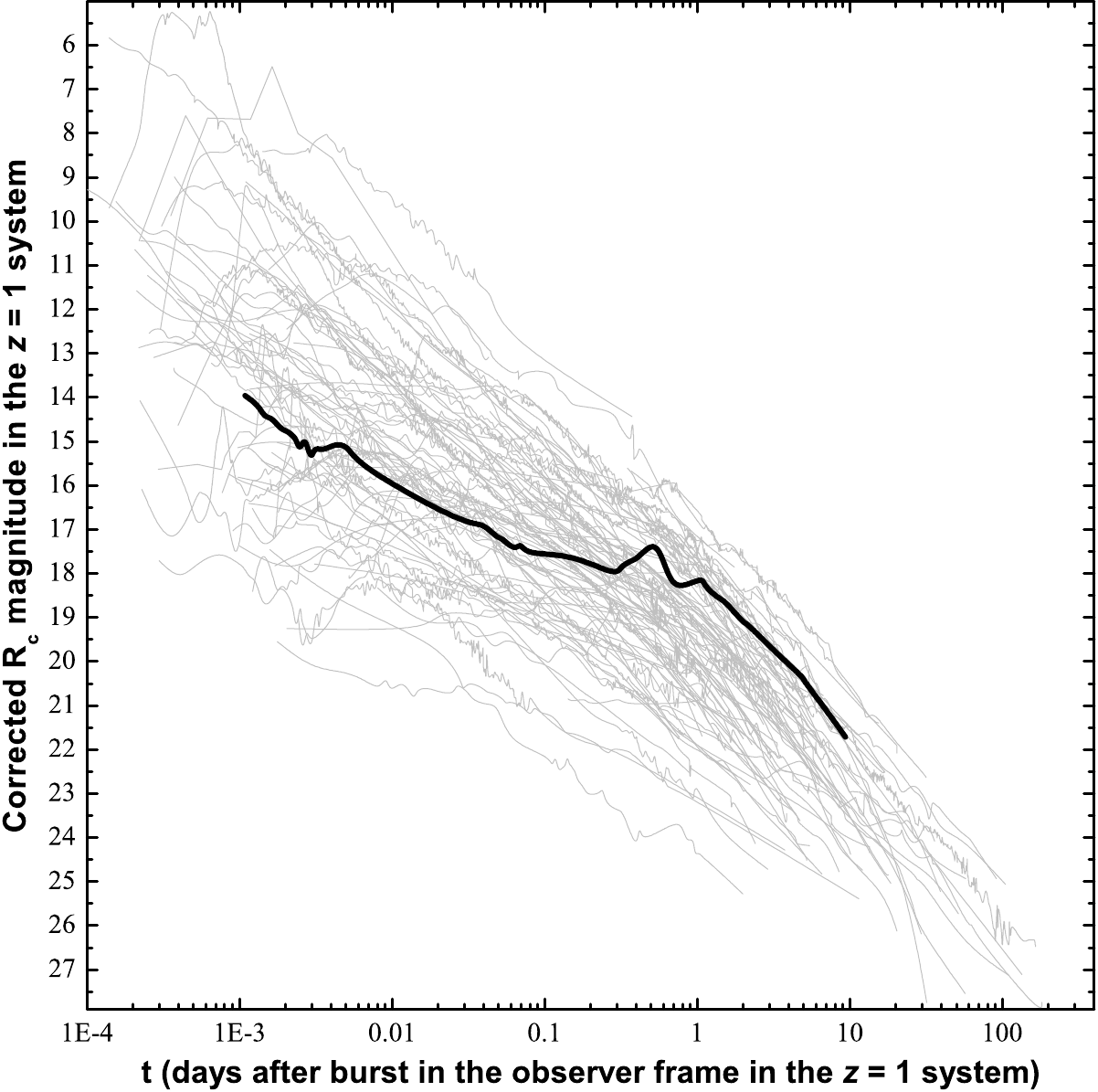}
\caption{The observed $R_C$-band afterglow of GRB\,110715A in comparison to a large sample of long GRB afterglows (left). After correction for the significant foreground extinction, it is seen to be one of the brightest afterglows ever observed. After correcting for rest-frame extinction and shifting to $z=1$ (right), the afterglow of GRB\,110715A is more common, although it remains among the more luminous detected to date at late times.}
\label{fig:complc}
\end{figure*}

\begin{figure}
\centering
\includegraphics[width=\columnwidth]{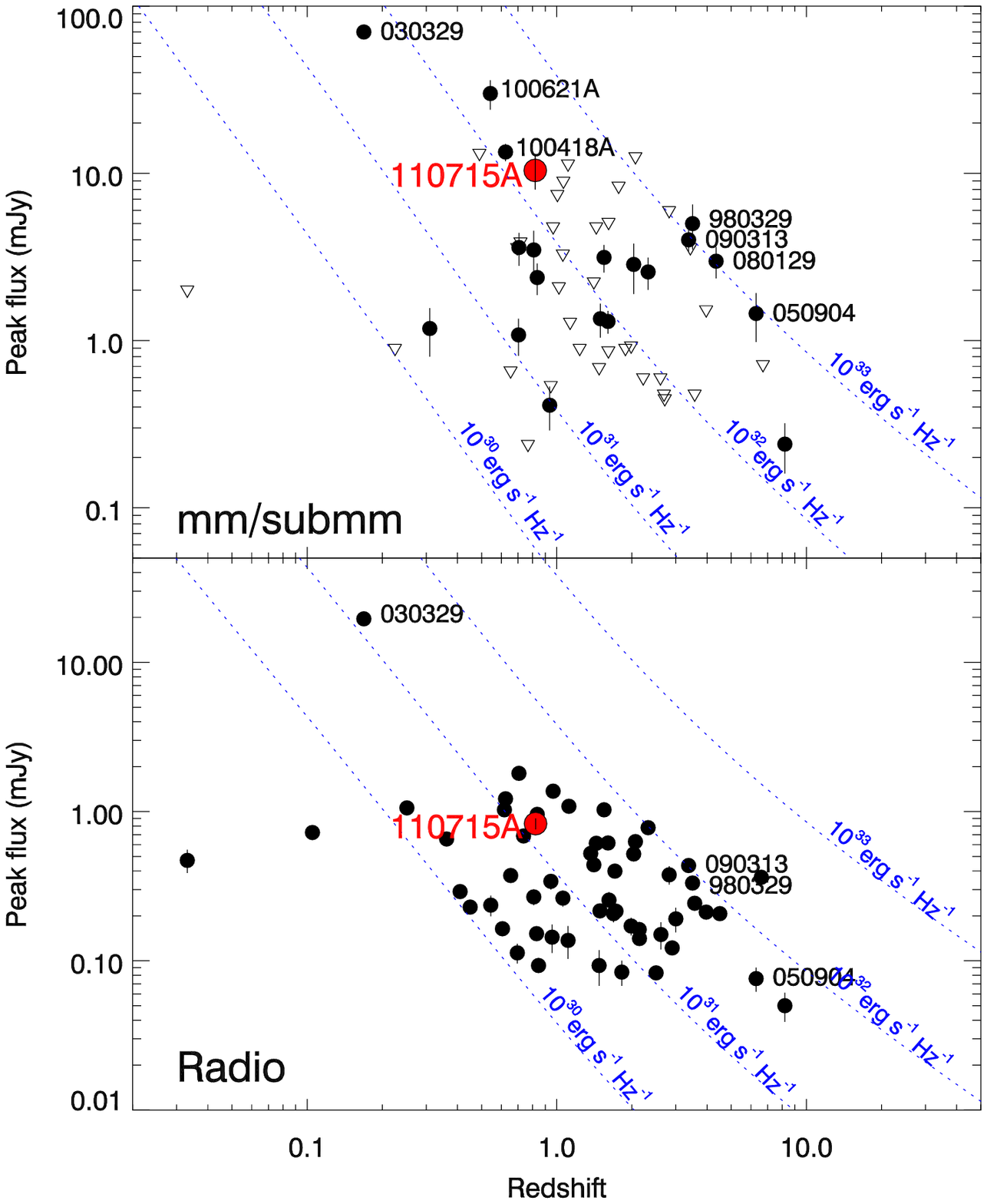}
\caption{Top: mm/submm afterglow as compared with the sample of \citet{2012A&A...538A..44D}. Bottom: Radio afterglow of GRB\,110715A compared with the sample of \citet{2012ApJ...746..156C}}
\label{fig:mmrad}
\end{figure}

Using the UVOT and GROND data, and adding the early $R_C$ band observations from \cite{2011GCN..12174...1N}, we construct a composite light curve by shifting all data to the $R_C$ band (no evidence for chromatic evolution is found). This light curve extends over almost four decades in time. Comparing it to the sample of long GRB afterglow light curves taken from \citet{2006ApJ...641..993K, 2010ApJ...720.1513K, 2011ApJ...734...96K}, we find that (after correcting for the significant -- 1.6 magnitudes -- foreground extinction) the afterglow is  among the brightest ever detected (especially after $\sim0.3$ days, see Fig.~\ref{fig:complc} left panel), comparable to those of GRB\,991208 \citep{2001A&A...370..398C} and GRB\,060729 \citep{2011MNRAS.413..669C}, both at lower redshift (see section \ref{sec:spec}). It becomes fainter than 20th magnitude only after about 4.5 days. Using the GROND data, we find a best fit for the Spectral Energy
Distribution (SED) of the afterglow with $\beta=0.90 \pm 0.22$, and a small (essentially zero) $A_V=0.09 \pm 0.18$ using SMC dust. With these data and knowledge of the redshift, we use the method of \citet{2006ApJ...641..993K} to shift the afterglow, corrected for all extinction, to $z=1$. We find a magnitude shift of $dRc=+0.38^{+0.17}_{-0.32}$. At one day after the trigger (in the $z=1$ frame), it is $R_C=17.97^{+0.19}_{-0.33}$, and $R_C=13.90^{+0.23}_{-0.35}$ at 0.001 days. This places the afterglow into the tight  peak found by \citet{2010ApJ...720.1513K} (their figure 6), which is formed by afterglows which are likely forward-shock dominated at early times already. This does not mean that a reverse shock component is not present. According to \citet{2010ApJ...720.1513K}, the early afterglow can be classified as ``Limit + Slow Decay'' \citep[][their  table 5]{2010ApJ...720.1513K}. In this sense, except for the rebrightenings, the afterglow is seen to be typical.

In Figure \ref{fig:mmrad}, we compare the radio and submm emission of GRB\,110715A to the samples of \citet{2012A&A...538A..44D} and \citet{2012ApJ...746..156C}: in submm, the afterglow peak brightness is among the brightest observed, with similar luminosity as GRB\,030329, GRB\,100621A or GRB\,100418A, but still an order of magnitude less luminous than the highest luminosity events (GRB\,980329, GRB\,090313, GRB\,080129 or GRB\,050904). The situation in radio is similar, with GRB\,110715A being amongst the brightest events.

At early times the physical size of the GRB afterglow emission region will be small, and thus there is the possibility of interstellar scintillation (ISS) modulating the observed flux of the afterglow, as has been seen previous in GRBs such as GRB\,970508 \citep{1997Natur.389..261F}. GRB\,970508 was seen to have large (40-50\%) fractional modulations in flux at 1.4 and 8.6~GHz for up to two months post burst, after which the intrinsic source size became large enough to break the conditions under which ISS is possible.

According to \citet{1998MNRAS.294..307W} the transition frequency at the location of GRB\,110715A is $\sim$ 40~GHz, indicating that, when present, scintillation should be strong (m$_d$ $>$ 1) below this frequency, and weak (m$_d$ $<$ 1) at higher frequencies. The temporal resolution of our observations during the first two weeks after the GRB is low so we are unable to measure scintillation. However, differences between the modelled and observed flux densities seen at $\lesssim$ 40~GHz in the first 2 weeks post burst are consistent with scintillation.

\subsection{Spectral absorption lines of the optical afterglow}
\label{sec:spec}

\begin{figure*}
\centering \includegraphics[width=0.8\textwidth]{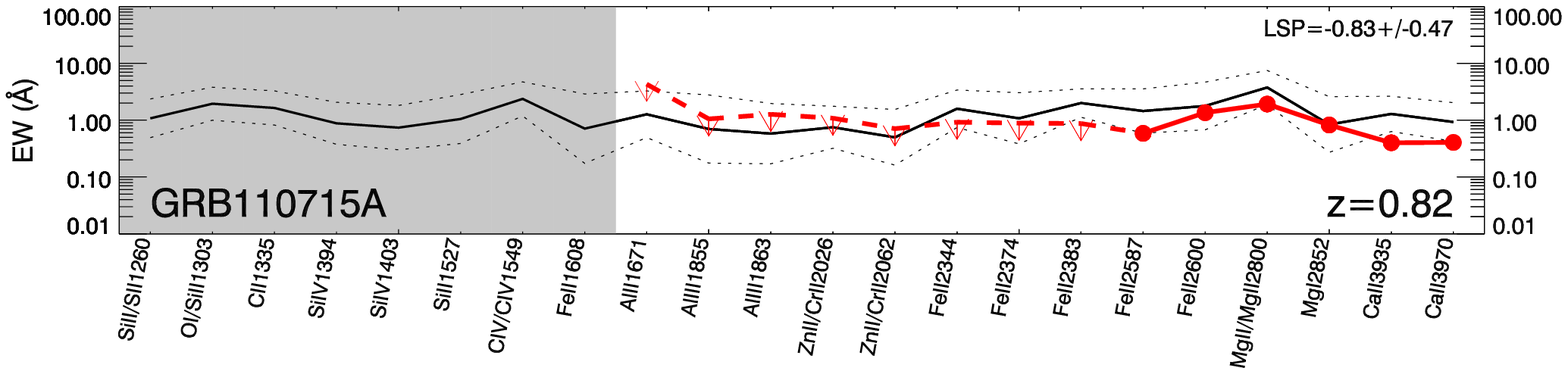}
\caption{Line strength diagram of the afterglow spectrum of GRB\,110715A, following the prescription of \citet{2012A&A...548A..11D}.}
\label{lsp}
\end{figure*}

We detect eight absorption features in the complete X-shooter spectrum that we identify as caused by \FeII, \MgII, \MgI, \CaII, and \CaI\ at a common redshift of 0.8225 $\pm$ 0.0001. For recent Planck cosmological parameters \citep[$\Omega_{\rm   M}$ = 0.315, $\Omega_\Lambda$ = 0.685, $H_0$ = 67.3 \kms\ Mpc$^{-1}$;][]{2014A&A...571A..16P}, this redshift corresponds to a luminosity distance of 5357.86 Mpc.

We have measured the equivalent widths of these lines and limits for several others using the self-developed code used in \citet{2009ApJS..185..526F} and \citet{2012A&A...548A..11D}. The results are shown in table \ref{table:ew}, as well as the composite afterglow spectrum by \citet{2011ApJ...727...73C} for comparison purposes. Using the prescriptions given by \citet{2012A&A...548A..11D}, we find that the neutral element population is higher than average. Detection of \CaI$\lambda 4227$, a line rarely observed in GRB afterglows \citep{2012A&A...548A..11D}, also supports the low ionisation hypothesis of the material in the line of sight to GRB\,110715A.

Following the prescription of \citet{2012A&A...548A..11D}, we obtain a line strength parameter for GRB\,110715A of LSP = -0.83 $\pm$ 0.47, implying that this event is in the percentile 13.4 of line strengths, and indicating a lower than average column density of material in the line of sight (86.6\% of GRBs have stronger lines). This often indicates a small host galaxy \citep{2012A&A...548A..11D}. This is consistent with the fact that there are no velocity components in the absorption features faster than 30 \kms.

\begin{table}

\caption{Features in the X-shooter spectra.}
\label{table:ew}
\begin{tabular}{lccc}
\hline\hline
Feature & $\lambda_{obs}$ [\AA] & $EW$ [\AA] & $EWc$ [\AA] $(a)$ \\
\hline
\AlII$\lambda$1671 & $\sim$3045 & $<$ 7.82 & 1.04 $\pm$ 0.02 \\
\AlIII$\lambda$1855 & $\sim$3380 & $<$ 1.92 & 0.89 $\pm$ 0.02 \\
\AlIII$\lambda$1863 & $\sim$3395 & $<$ 2.32 & 0.68 $\pm$ 0.02 \\
\ZnII$\lambda$2026+\CrII$\lambda$2026 & $\sim$3692 & $<$ 1.97 & 0.60 $\pm$ 0.02 \\
\CrII$\lambda$2062+\ZnII$\lambda$2063 & $\sim$3758 & $<$ 1.30 & 0.53 $\pm$ 0.02 \\
\FeII$\lambda$2261 & $\sim$4120 & $<$ 1.81 & 0.38 $\pm$ 0.02 \\
\FeII$\lambda$2344 & $\sim$4272 & $<$ 1.68 & 1.74 $\pm$ 0.02 \\
\FeII$\lambda$2374 & $\sim$4327 & $<$ 1.63 & 1.00 $\pm$ 0.02 \\
\FeII$\lambda$2383 & $\sim$4342 & $<$ 1.60 & 1.65 $\pm$ 0.02 \\
\FeII$\lambda$2587 & 4714.04 & 1.07 $\pm$ 0.47 & 1.33 $\pm$ 0.02 \\
\FeII$\lambda$2600 & 4737.32 & 2.47 $\pm$ 0.73 & 1.85 $\pm$ 0.03 \\
\MgII$\lambda$2796 & 5096.44 & 1.99 $\pm$ 0.34 & 1.71 $\pm$ 0.02 \\
\MgII$\lambda$2803 & 5109.09 & 1.50 $\pm$ 0.31 & 1.47 $\pm$ 0.02 \\
\MgI$\lambda$2853 & 5198.29 & 1.50 $\pm$ 0.40 & 0.78 $\pm$ 0.01 \\
\CaII$\lambda$3935& 7171.01 & 0.72 $\pm$ 0.07 & 0.76 $\pm$ 0.02 \\
\CaII$\lambda$3970 & 7234.51 & 0.74 $\pm$ 0.07 & 0.66 $\pm$ 0.02 \\
\CaI$\lambda$4228 & 7705.26 & 0.37 $\pm$ 0.06 & 0.11 $\pm$ 0.02 \\
\hline
\end{tabular}
\\[0.2cm]
\begin{scriptsize}
(a) Equivalents widths measured on the composite GRB afterglow spectrum \citep{2011ApJ...727...73C}.
\end{scriptsize}
\end{table}

\subsection{The host galaxy}

We computed the distance between the afterglow and the host galaxy core. The centroid is offset by 0.21 $\pm$ 0.03 \arcsec\ with respect to the ALMA position, which at the redshift of GRB\,110715A corresponds to 1.56 $\pm$ 0.19 kpc. This is comparable to the typical offset of 1.2 kpc seen for long GRBs \citep{2002AJ....123.1111B,2012A&A...548A..11D}. The host absolute magnitude (AB) would be $M$ = -18.2 mag at a restframe wavelength of 4200 \AA, which is similar to the Johnson $B$-band (without needing to make assumptions on the host galaxy spectral index). 

The luminosity of the host galaxy is low, even relative to other GRB hosts \citep[which tend to occur in lower-mass and lower-luminosity galaxies than average at $z \lesssim 1.5$; e.g.][]{2016ApJ...817....7P}, although it is by no means extreme or exceptional. For example, relative to the UV luminosity distribution of nine galaxies at roughly similar redshift ($0.5<z<1.1$) in the TOUGH sample \citep{2015ApJ...808...73S}, this host galaxy is less luminous than six or more, depending on the unknown $k$-correction across the Balmer break which is not known for the TOUGH sample. We also compared this magnitude to synthetic $B$-band magnitudes of galaxies from the larger, multi-colour SHOALS sample \citep[][and work in prep.]{2016ApJ...817....8P}. The host of GRB\,110715A is about 0.6 mag less luminous than the median $B$ magnitude of $0.5<z<1.1$ galaxies in this sample, and is more luminous than only five out of these 21 galaxies. Compared to a more local galaxy population, it is slightly more luminous than the LMC ($M_B \sim -17.5$) but of course much less luminous than nearby spirals such as the Milky Way or M31 ($M_B \sim -20.5$ to $-21$). This faint host galaxy is consistent with the faint and unresolved absorption features seen in the afterglow spectrum. Considering also its very low ionisation environment, all evidence suggest that the sight-line towards GRB\,110715A is probing an small dwarf host galaxy, maybe in its initial star-forming episode due to the low background ionising radiation, which keeps an unusual abundance of \CaI.

\subsection{Modeling of the afterglow evolution}
\label{sec:agm}

\subsubsection{Model and fitting description}

\begin{table}
\caption{The lower and upper boundaries of the priors on parameters used in the analysis.}
\label{tab:priors}
\begin{tabular}{llll}
\hline\hline
Parameter & Distribution & Lower & Upper \\
\hline
$E_{iso}$ [$10^{53}$ erg] & log-uniform & 0.01 & 10000 \\
$\Gamma_0$ & log-uniform & 10 & 2000 \\
$\theta_0$ [deg] & log-uniform & 0.1 & 90 \\
$p$ & uniform & 1.1 & 4.0 \\
$\epsilon_i$ & log-uniform & 0.0001 & 0.5 \\
$\epsilon_e$ & log-uniform & 0.0001 & 0.5 \\
$\epsilon_B$ & log-uniform & 0.0001 & 0.5 \\
$A_*$ [$5.015 \cdot 10^{11}$ cm$^{-3}$] & log-uniform & 0.00001 & 100 \\
$n_0$ [cm$^{-3}$] & log-uniform & 0.0001 & 1000 \\
$t_{sh}$ [min] & log-uniform & 0.00001 & 200 \\
$r_{sh}$ & uniform & 1 & 50 \\
$t_1^{(a)}$ [days] & log-uniform & 0.0001 & 200 \\
$E_1/E_0^{(b)}$ & uniform & 0 & 50 \\
$A_{V, host}$ [mag] & uniform & 0 & 1.0 \\
$E(B-V)$ [mag] & Gaussian$^{(c)}$ &  & \\
\hline
\end{tabular}
\\[0.2cm]
\begin{scriptsize}
(a) $t_1$ is the time in the observer's frame at which the energy injection catches up with the forward shock. \\
(b) $E_1$ is the energy of the injection and $E_0$ is the initial energy release. \\
(c) $\mu$ = 0.56, $\sigma$ = 0.04
\end{scriptsize}
\end{table}

The afterglow emission was modeled with the numerical code of \citet{2006ApJ...647.1238J}. This software has been used successfully to model several different afterglows, including GRB\,060121 \citep{2006ApJ...648L..83D}, GRB\,050408 \citep{2007A&A...462L..57D}, GRB\,060526 \citep{2010A&A...523A..70T}, and GRB\,050525A \citep{2012MNRAS.427..288R}. This model assumes the emission originates in a forward shock only, with a top-hat jet configuration. The algorithm simulates that a slab of matter with mass $M_0$ is ejected with a Lorentz factor of $\Gamma_0$ into a cone with a half-opening angle of $\theta_0$. The slab starts accumulating matter and slows down in the process. Energy injections ($E_i$) at a time $t_i$ are modelled as slabs of matter moving at lower speeds than the forward shock ($\Gamma_i < \Gamma_0$) and catching up to it at later times. At the time of collision, the energy and momentum of the forward shock of the injected slab is instantaneously added to the already moving mass. The emission from any reverse shock formed in the collision is ignored. To calculate the emission, we assume that a fixed fraction of the energy of the forward shock is contained in the magnetic field and electron distribution of the forward shock. For the magnetic field, this fraction is denoted with $\epsilon_B$. In \citet{2006ApJ...647.1238J}, the fraction of energy contained in the electrons was denoted with $\epsilon_e$. This is now  changed, to allow for the slope of the electron power-law distribution, $p$, to be less than 2. We used the formalism of \citet{2001ApJ...554..667P} and denote with $\epsilon_i$ the fraction of energy contained in the electrons with the lowest energy in the distribution. The highest energy in the distribution is then limited such that the total energy of the electron distribution never exceeds a fraction $\epsilon_e$ of the forward shock energy.

To explain the data, we need a model that includes a temporary increase in flux around 0.3 days after the onset of the GRB that is observed in the light curves shown in Figure~\ref{fig:obs}. We chose three different types of models that we expect have this behaviour: a model with a constant density interstellar medium ($n_0$) and a single energy injection (CM), a model with a wind density external medium ($\rho = A_* r^{-2}$) and a single energy injection (WM), and a model with a wind termination shock (with fractional change in density at the shock front denoted by $r_{sh}$) but no energy injection (TS).

The best fit model parameters are found using Bayesian inference using the MultiNest tool \citep{2009MNRAS.398.1601F}. MultiNest is well suited for exploring the parameter space of the non-linear afterglow model and finds parameter correlation as well as all modes in the parameter space fitting the data similarly well. In addition to the afterglow model parameters, we also determine the host dust extinction in the fit, which we assume follows an SMC-like extinction curve. It is also possible to let the Galactic dust extinction vary as a nuisance parameter. This is of special interest in our case due to the large and uncertain value along the GRB line of sight through our Galaxy.

One of the main benefits of a Bayesian analysis is the introduction of prior distributions on parameters. For this analysis we have unfortunately very little prior knowledge on their values. We therefore opted for flat priors on all parameters, but Galactic reddening, and made sure the parameter limits were large enough so that the posterior is not affected by these limits unless they are physical (see Table~\ref{tab:priors}). Examples of such physical boundaries are the requirements that the extinction of the host be positive ($A_{V, host} > 0$) and the fractional change in density at the shock front should not decrease ($r_{sh} > 1$). We also constrain the fraction of energy in the electrons ($\epsilon_e$) and magnetic field ($\epsilon_B$) such that the fraction of energy contained in the rest of the jet, $\epsilon = 1 - \epsilon_e - \epsilon_B$, is larger than both $\epsilon_e$ and $\epsilon_B$.  This is to make sure the jet's energy is not dominated by that of the electrons and the magnetic field. The constraint is not hard and $\epsilon$ is usually somewhere in between $\epsilon_e$ and $\epsilon_B$ if both are large like in this analysis. There is also the hard prior that $\epsilon_e > 1.1 \epsilon_i$ so the energy in the total electron distribution is always at least 10 percent greater than that contained in the electrons with the lowest value. This constraint is actually reached in all of our models, resulting in a strong correlation between $\epsilon_e$ and $\epsilon_i$ (see Fig.~\ref{fig:marginal2d}).

Due to several reasons, such as the high Galactic reddening, the wavelength range on which \HI\ absorption is located, and the difficulties to compute the effective wavelength of the UVOT \textit{white} filter due to its band width, we performed different fits in order to identify and quantify the sources of systematic uncertainties. We excluded the upper limits from the UV filters of UVOT as well as observations using its $u$-band filter because of the uncertainty on the Galactic dust extinction (none of  the models are able to accurately reproduce those data points, either when included in the fit or not). Below we discuss the best fit results that were obtained, and refer the interested reader to the material contained in the Appendix \ref{app:view} and online material for the result details of the complete set of Bayesian fits.

\subsubsection{The best fit}

\begin{figure*}
\centering
\includegraphics[width=\textwidth]{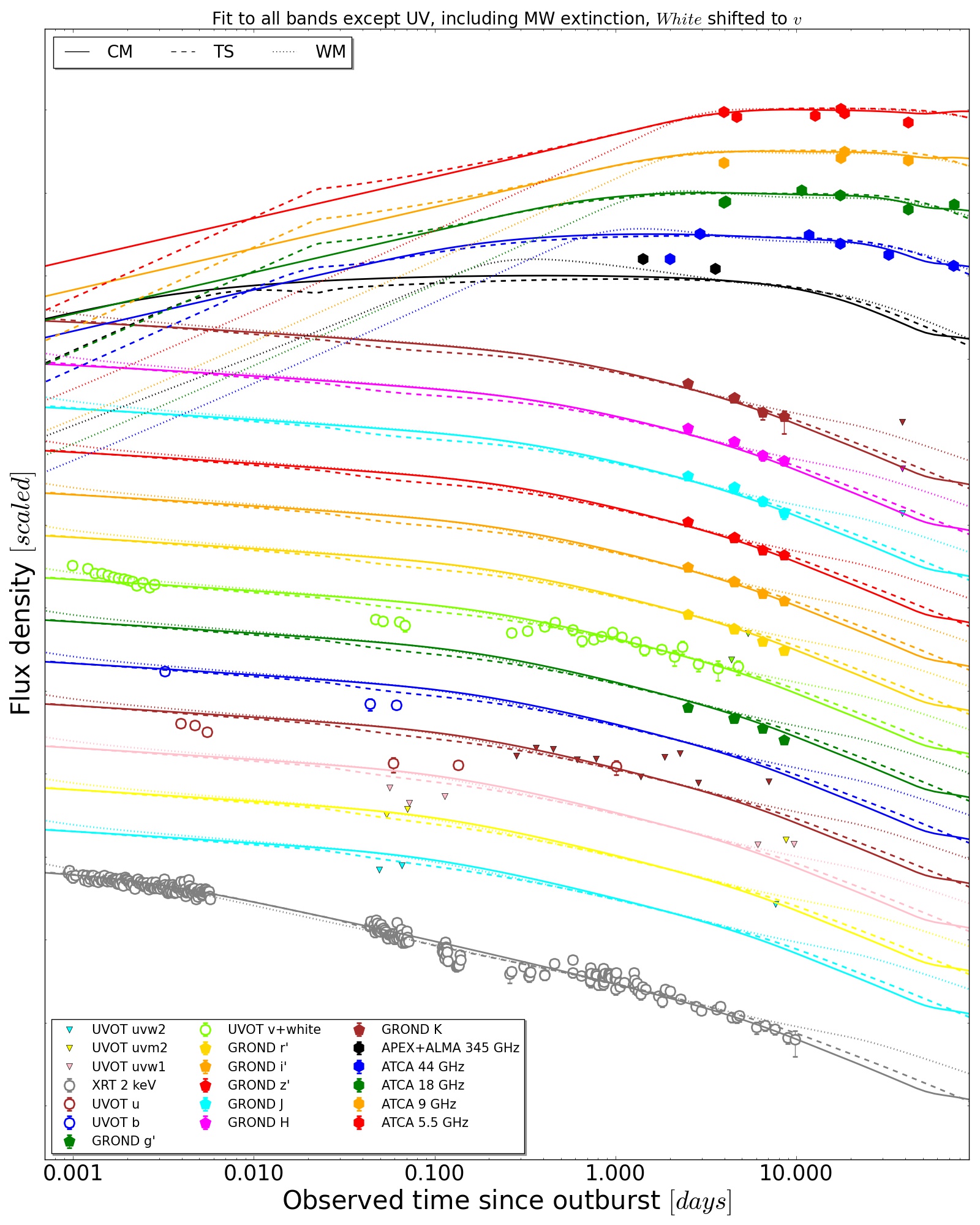}
\caption{Best fits to the GRB\,110715A light curves. Fluxes are independently scaled for each band for clarity. The full set of plots for each subset is available in the online version.}
\label{fig:lc}
\end{figure*}

\begin{figure*}
\centering
\includegraphics[width=\textwidth]{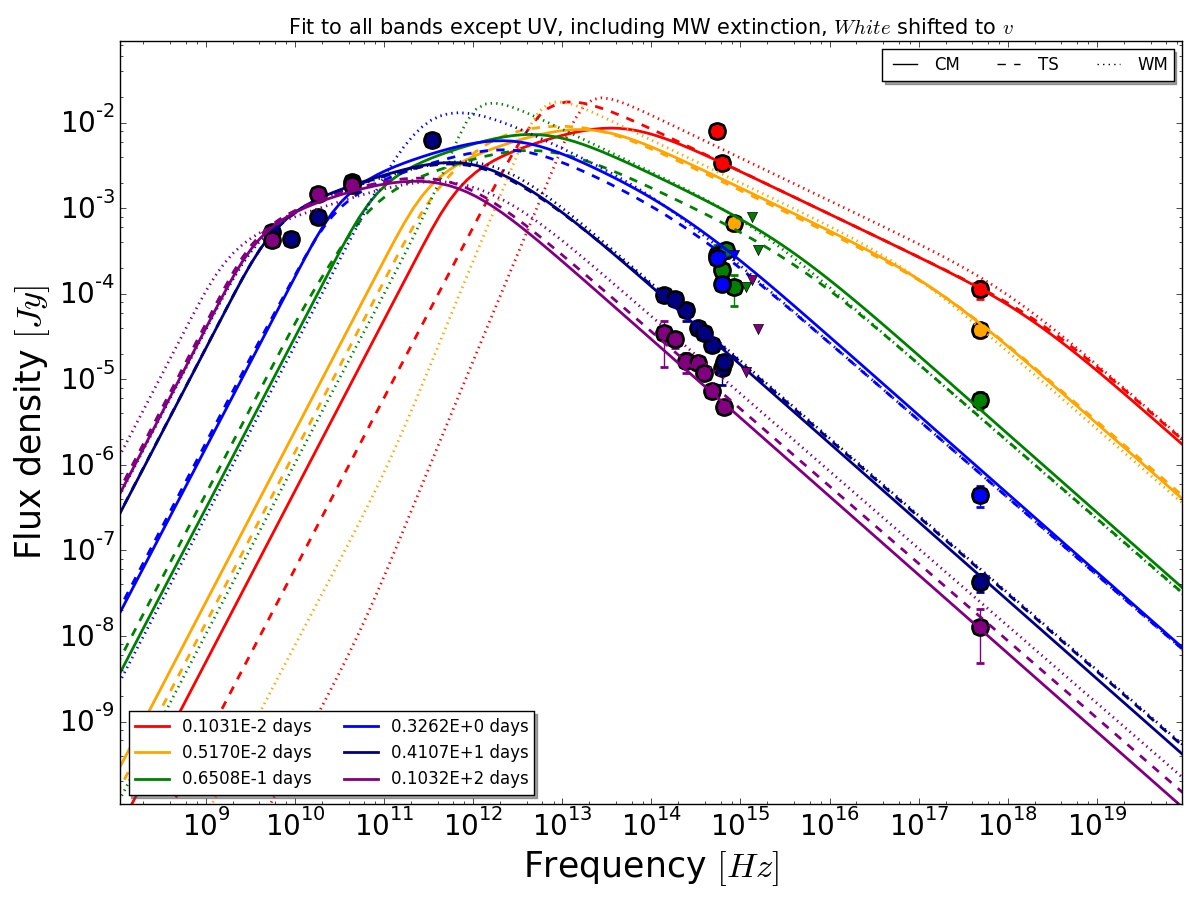}
\caption{Best fits to the SED for GRB\,110715A observed at several epochs. The full set of plots for each data subset is available in the online version.}
\label{fig:sed}
\end{figure*}

\begin{figure*}
\centering
\includegraphics[width=\textwidth]{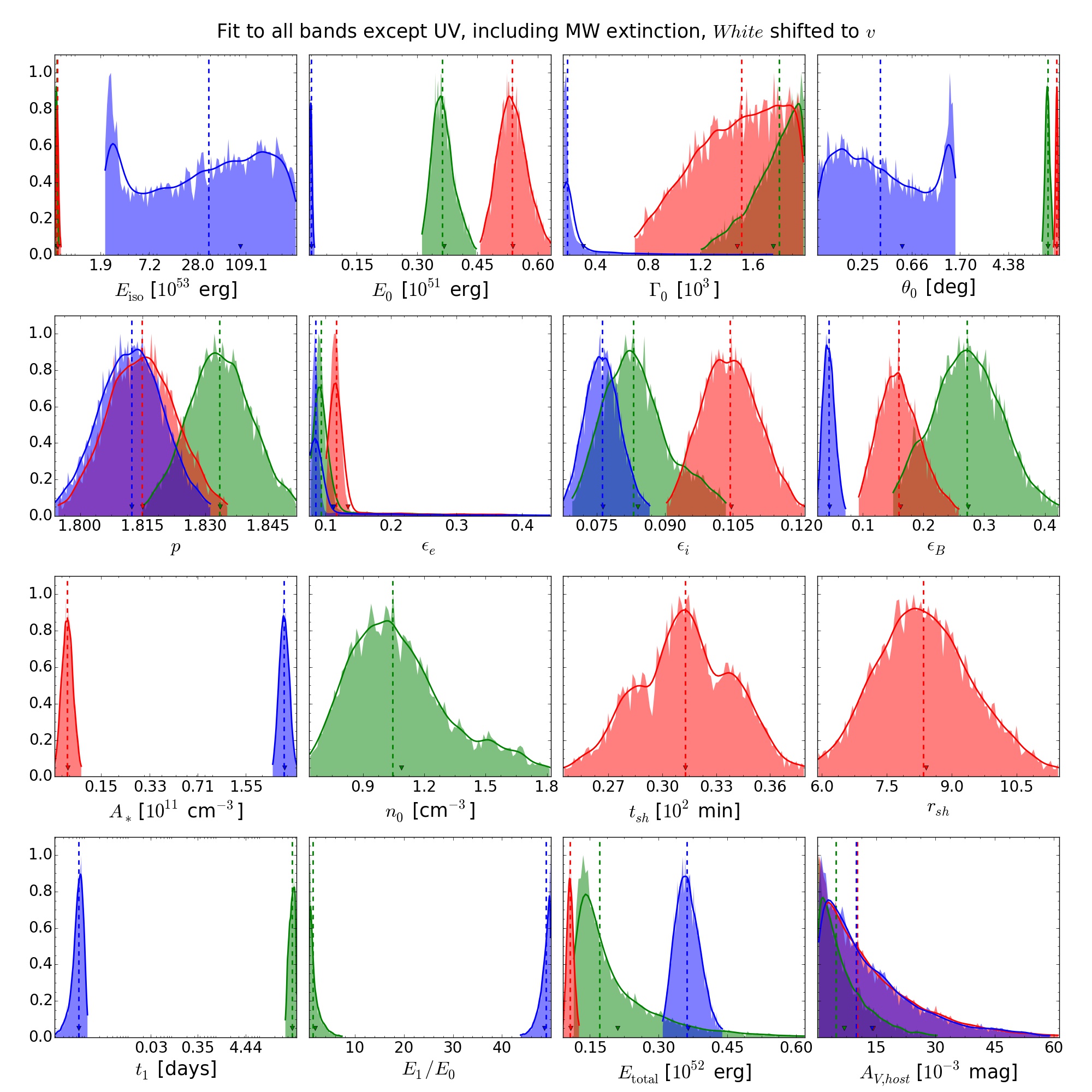}
\caption{Posterior distributions of the three models for the best fit data subset. In all marginal plots, CM is represented in green, TS in red, and WM in blue. The full set of plots for each set is available in the online version.}
\label{fig:marginal1d}
\end{figure*}

\begin{figure*}
\centering
\includegraphics[width=\textwidth]{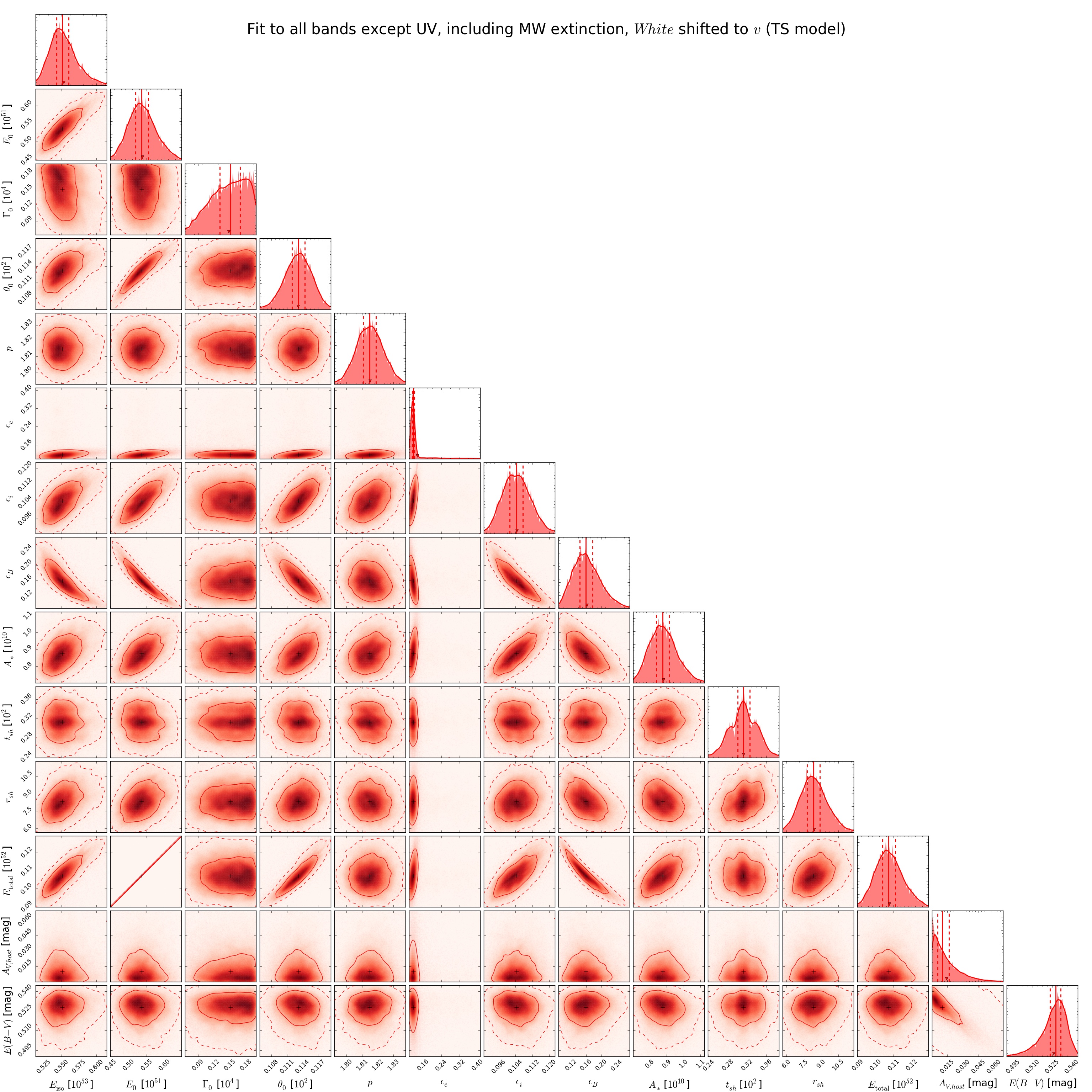}
\caption{Parameter correlations for the best fit model. The full set of plots for each data subset is available in the online version.}
\label{fig:marginal2d}
\end{figure*}

\begin{table*}
\caption{The Bayesian evidence and the parameter posterior mean as reported by MultiNest for the three different models of the best fit data-set. The full table for each data subset is available in the online version.}
\label{tab:pars}
\begin{tabular}{l|ccc}
Parameter & CM & TS & WM \\
\hline\hline
Evidence & --1015.83 & --995.70 & --1129.24 \\
$\chi^2_r$ & 5.73 & 5.50 & 7.30 \\
$\chi^2_{r, x}$ $^{(a)}$ & 2.47 & 2.15 & 2.76 \\
$\chi^2_{r, o}$ $^{(b)}$ & 13.00 & 11.34 & 17.48 \\
$\chi^2_{r, r}$ $^{(c)}$ & 26.79 & 32.54 & 36.38 \\
$E_{\rm iso}$ [erg] & $53.63^{+0.66}_{-0.62} \times 10^{51}$ & $55.10^{+0.92}_{-0.82} \times 10^{51}$ & $0.38^{+0.60}_{-0.26} \times 10^{55}$ \\
$E_0$ [erg] & $3.63^{+0.14}_{-0.12} \times 10^{50}$ & $5.36^{+0.18}_{-0.17} \times 10^{50}$ & $3.64^{+0.15}_{-0.14} \times 10^{49}$ \\
$\Gamma_0$ & $1799^{+82}_{-110}$ & $1510^{+180}_{-200}$ & $184^{+35}_{-12}$ \\
$\theta_0$ [deg] & $9.44^{+0.22}_{-0.20}$ & $11.32 \pm 0.12$ & $0.35^{+0.27}_{-0.13}$ \\
$p$ & $1.8334^{+0.0038}_{-0.0036}$ & $1.8148 \pm 0.0041$ & $1.8124^{+0.0037}_{-0.0039}$ \\
$\epsilon_e$ & $9.32^{+0.57}_{-0.41} \times 10^{-2}$ & $(11.64 \pm 0.40) \times 10^{-2}$ & $(8.53 \pm 0.26) \times 10^{-2}$ \\
$\epsilon_i$ & $(8.31 \pm 0.31) \times 10^{-2}$ & $(10.44 \pm 0.32) \times 10^{-2}$ & $(7.62 \pm 0.19) \times 10^{-2}$ \\
$\epsilon_B$ & $(2.72 \pm 0.28) \times 10^{-1}$ & $(1.59 \pm 0.16) \times 10^{-1}$ &  $(4.44 \pm 0.47) \times 10^{-2}$ \\
$A_*$ [$5.015 \cdot 10^{11}$ cm$^{-3}$] &  & $0.01747^{+0.00078}_{-0.00074}$ & $0.571^{+0.023}_{-0.022}$ \\
$n_0$ [cm$^{-3}$] & $1.05^{+0.12}_{-0.10}$ &  &  \\
$t_{sh}$ [min] &  & $3.13^{+0.13}_{-0.12} \times 10^{1}$ &  \\
$r_{sh}$ &   & $8.33^{+0.56}_{-0.54}$ &  \\
$t_1$ [days] & $5.03^{+0.42}_{-0.44} \times 10^{1}$ &   & $(6.79 \pm 0.91) \times 10^{-4}$ \\
$E_1/E_0$ & $1.34^{+0.59}_{-0.35}$ &   & $49.01^{+0.43}_{-0.61}$ \\
$E_{\rm total}$ [erg] & $1.72^{+0.42}_{-0.24} \times 10^{51}$ & $10.72^{+0.36}_{-0.34} \times 10^{50}$ & $3.62^{+0.14}_{-0.12} \times 10^{51}$ \\
$A_{V, host}$ [mag] & $0.0048^{+0.0031}_{-0.0021}$ & $0.0102^{+0.0063}_{-0.0046}$ & $0.0099^{+0.0064}_{-0.0042}$ \\
$E(B-V)$ [mag] & $0.5249 \pm 0.0030$ & $0.5277^{+0.0037}_{-0.0044}$ & $0.5749^{+0.0035}_{-0.0043}$ \\
\hline
\end{tabular}
\\[0.2cm]
\begin{scriptsize}
(a) $\chi^2$ computed only with the X-ray data.\\
(b) $\chi^2$ computed only with the UV, optical and nIR data.\\
(c) $\chi^2$ computed only with the submm and mm data.\\
\end{scriptsize}
\end{table*}

\begin{figure}
\centering
\includegraphics[width=\columnwidth]{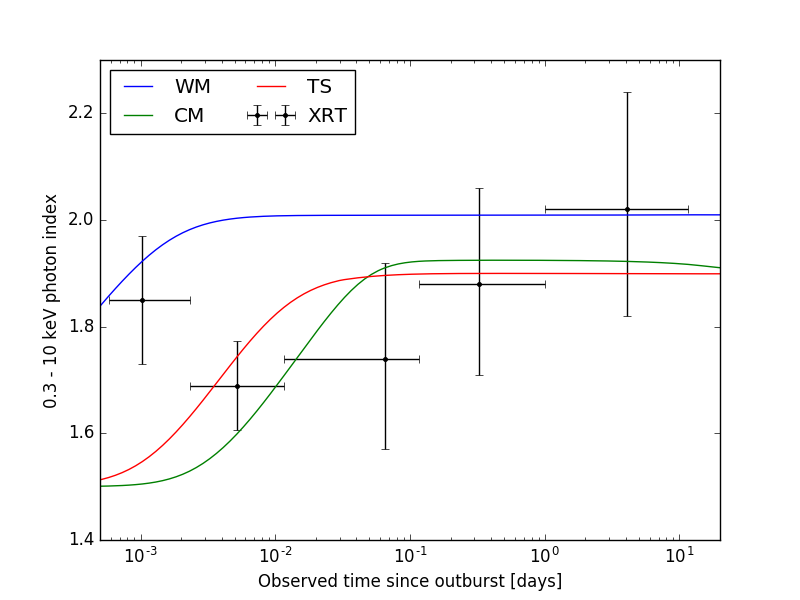}
\caption{Best fit of the evolution of the photon index to that found from XRT data analysis. The full set of plots for each subset is available in the online version.}
\label{fig:photon}
\end{figure}

We found the best fit models to be those in which we excluded the UV bands, we set MW extinction as a nuisance parameter, and the UVOT \textit{white} band was shifted to V (see Table~\ref{tab:goodness} for a detailed statistical analysis of the results for the complete grid of fits). This table suggests the TS fit as the most plausible model that describes the afterglow. However, none of them is either a statistically good fit or can fully explain the temporary flux increase at $\sim$0.3~days. We can also observe that there are no statistical arguments to reject most of the fits when compared with the best. Fit results are shown overlaid on the data in Figures \ref{fig:lc} (light curves) and \ref{fig:sed} (spectral energy distributions).

The time of the wind termination shock for the best TS model agrees very well with the time of the flux increase in the light curve. The density increases by a factor of 8.3 at that time, more than 2 times the expected density increase for strong shocks. Despite this, the effect on the light curves is not strong enough to explain the data. Due to the spectral parameters required in the fit, the wind termination shock causes a flux decrease rather than an increase as the cooling break is just below the optical band. The slow decay in the early X-ray light curve in the model is caused by the injection break being above the X-ray frequency. This requires there to be a spectral evolution in the X-ray light curve that is not observed. In both cases the preferred location of the energy injection is at a different point than expected. For the best fit of the CM model it is much later and serves only to explain the latest radio points while for the WM model it happens very early to explain the shallower decay between 0.01 and 1 days. The best model also has a hard time explaining the rapid decline in the light curve observed by the GROND instrument. The earliest points are under-predicted while the later points are over-predicted. This is again something that all the models fail to reproduce. The CM model does a slightly better job, but the WM model is worst. The models have a similarly hard time at explaining the late time X-ray light curve as they don't decay rapidly enough.

Figure \ref{fig:marginal1d} compares the evolution of the photon index of our best fit models to that found from XRT data analysis. The CM and TS models do a reasonable job of explaining the X-ray spectral evolution with the exception of the first time bin where the models are harder than observed. The WM model agrees with the first time bin but is then too soft for the next two bins.

Finally, the best model is unable to explain the early and late radio and submm data. The model has a hard time explaining the rapid rise of the 44~GHz data simultaneously with the decay of the 345~GHz data and the rather flat light curve at 5.5~GHz. The self-absorption break, $\nu_a$, needs to pass through the 44~GHz band at around 2 days to explain the rapid increase and it should have already passed the 345~GHz band at 1 day and the 5.5~GHz band at 2 days. It is impossible for the model to meet these criteria. In addition, the 44~GHz light curve starts decaying at around 10 days with a slope that is incompatible with post jet-break evolution and at the same time the 18~GHz band is compatible with being nearly constant. The CM model has similar issues as the TS model although it does slightly better at late times because the injection lifts the radio light curve to match the last points. There is, however, no other indication for the energy injection and it is unlikely to be the correct physical interpretation. The WM model does the best job with the submm and radio data, but is still far from explaining the details of the observed afterglow.

Table~\ref{tab:pars} shows the best fit model posterior median for the parameter values and their associated 68\% statistical errors. Figure~\ref{fig:marginal1d} shows a plot comparing the resulting distributions of the models for the best data subset. The parameter values that give the smallest $\chi^2$ are usually located near the peak of the posterior distributions, and their distributions are mostly symmetric, with notable exceptions in the WM model where long tails can be seen for $E_{\rm iso}$, $\Gamma_0$, and $\theta_0$. The inferred values are mostly typical for a GRB afterglow with a few exceptions. The half opening angle $\theta_0$ in the WM model is unreasonably low and requires an extreme confinement of the outflow. It is also the worst performer of the 3 models and we therefore consider it an unreasonable model. The value of the electron power-law index $p$ is on the lower side and lower than expected from shock-acceleration theory \citep{2001MNRAS.328..393A}. It is, however, well within the range of values deduced from observations of relativistic shocks \citep{2006MNRAS.371.1441S}. The fraction of energy contained in the magnetic field and the electron distribution, $\epsilon_B$, and $\epsilon_e$, respectively is rather large. $\epsilon_i$ is also larger than usual and constrained mostly by the equipartition requirements for the electron energy distribution rather than the position of the injection peak in the synchrotron spectrum. These large values cause the afterglow to be in the fast-cooling regime for the entire duration of the afterglow and the assumption of no radiative losses is likely invalid.

Our values for the host extinction are compatible with the value being very small as found earlier using GROND data only. The statistical error is significantly smaller because we use  the entire data-set, but the exact value is model dependent. As already discussed, the Galactic line of sight extinction is important in the GRB direction. The expected uncertainty could be large too, so similarly to the host galaxy, the Galactic $E(B-V)$ was left in some fits as a free parameter. Resulting Galactic and host galaxy solutions show a clear anti-correlation, limiting our constraints of the inferred host galaxy A$_{V, host}$. In Figure~\ref{fig:marginal2d} we include the two dimensional marginal plots for the $E(B-V)$ of the Galactic extinction with the rest of the parameters when it is included in the fit. As shown in Appendix~\ref{app:view}, the galactic $E(B-V)$ value is model dependent, and results are somewhat bimodal, sometimes consistent with the dust maps of \citet{1998ApJ...500..525S}, and sometimes with \citet{2011ApJ...737..103S}, the latter being the more favoured. The figures also show that the upper value found for the $E(B-V)$ parameter in this analysis is basically bound by the host extinction going to 0. 

We also discuss in Appendix \ref{app:view} the hypothesis that the reverse shock (RS) contribution has to be taken into account \cite[e.g.,][]{2006A&A...454L.119J, 2013ApJ...776..119L}, as well as the possibility that a double jet model \citep[e.g.][]{2005MNRAS.360..305S, 2008Natur.455..183R, 2011A&A...526A.113F, 2014MNRAS.444.3151V} is necessary to explain this afterglow. The conclusion is that, in spite of them possibly being present, none of the options can improve the fit starting at 0.5 days after the trigger, so other considerations must be taken into account to improve the fits we performed.

\section{Conclusions}

We present an extensive follow-up of the afterglow of GRB\,110715A in 17 bands ranging from a few seconds up to 74 days after the trigger. The line of sight is affected by strong foreground Galactic extinction, which complicated the follow-up and the analysis of the data.

GRB\,110715A had a very bright afterglow at all wavelengths, although its intrinsic luminosity is not exceptional.

Optical/nIR spectroscopy obtained with X-shooter shows weak absorption features at a redshift of $z$ = 0.8224 with no resolved velocity components ($\lesssim$ 30 \kms). Absorption line ratios indicate a low ionization environment, confirmed by the rare detection of \CaI.

Deep late imaging reveals a faint host galaxy with an absolute magnitude of M$_B$ = -18.2. This is consistent with the weak absorption features detected in the spectrum.

We attempted to model the broadband data with a fireball model based on the prescription of \citet{2006ApJ...647.1238J}. The best model implies a forward shock evolving through a wind environment with a termination shock. In spite of describing roughly the behavior of the afterglow, none of the models is able to get a statistically acceptable fit. This shows the need for better broadband sampling and more complex models to accurately describe the physics of GRB afterglows. There are several works that explore other possibilities, such as magneto-hydrodynamic simulations \citep{2012ApJ...749...44V}, which was satisfactorily used, e.g., in \citet{2014MNRAS.438..752G,2015ApJ...799....3R,2015ApJ...806...15Z}, or central engine activities \citep{2014ApJ...787...66Z}. These and other effects might be considered together in future works to get a more accurate view of the GRB afterglow physics.

Radio and sub-mm, along with X-ray observations, have been proven to be the most constraining bands for the afterglow modeling. We were limited by sensitivity for a long time in the crucial wavelength range of sub-mm, but now that ALMA is available, we have a good chance of getting high-quality data for a larger number of GRBs. This new, current and future facilities will allow us to probe the emission mechanisms in greater detail than previously possible, and will be determinant in the evolution of the GRB afterglow models.

\section*{Acknowledgements}

In memory of Javier Gorosabel: an exceptional supervisor, a brilliant scientist, and an even better human being.
RSR is grateful to SEPE for financial support while finishing this work and his PhD thesis.
RSR, SRO, AJCT, YDH, SJ, and JCT acknowledge financial support of the Spanish Government projects AYA 2009-14000-C03-01 and AYA 2012-39727-C03-01.
Parts of this research were conducted by the Australian Research Council Centre of Excellence for All-sky Astrophysics (CAASTRO), through project number CE110001020.
AdUP and CT acknowledge support from Ram\'{o}n y Cajal fellowships and from the Spanish research project AYA 2014-58381.
JJ acknowledges financial contribution from the grant PRIN MIUR 2012 201278X4FL 002 "The Intergalactic Medium as a probe of the growth of cosmic structures".
DAK acknowledges financial support by MPE Garching and TLS Tautenburg.
Part of the funding for GROND (both hardware as well as personnel) was generously granted from the Leibniz-Prize to Prof. G. Hasinger (DFG grant HA 1850/28-1).
PS and TK acknowledges support through the Sofja Kovalevskaja Award to P. Schady from the Alexander von Humboldt Foundation of Germany.
AU is grateful for travel funding support through the Max-Planck Inst. for Extraterrestrial Physics.
SK and ANG acknowledge support by DFG grant Kl 766/16-1.
ALMA is a partnership of ESO (representing its member states), NSF (USA) and NINS (Japan), together with NRC (Canada), NSC and ASIAA (Taiwan), and KASI (Republic of Korea), in cooperation with the Republic of Chile. The Joint ALMA Observatory is operated by ESO, AUI/NRAO and NAOJ.
This work made use of data supplied by the UK Swift Science Data Centre at the University of Leicester.




\bibliographystyle{mnras}
\bibliography{grb} 

\begin{thebibliography}{}
\makeatletter
\relax
\def\mn@urlcharsother{\let\do\@makeother \do\$\do\&\do\#\do\^\do\_\do\%\do\~}
\def\mn@doi{\begingroup\mn@urlcharsother \@ifnextchar [ {\mn@doi@}
  {\mn@doi@[]}}
\def\mn@doi@[#1]#2{\def\@tempa{#1}\ifx\@tempa\@empty \href
  {http://dx.doi.org/#2} {doi:#2}\else \href {http://dx.doi.org/#2} {#1}\fi
  \endgroup}
\def\mn@eprint#1#2{\mn@eprint@#1:#2::\@nil}
\def\mn@eprint@arXiv#1{\href {http://arxiv.org/abs/#1} {{\tt arXiv:#1}}}
\def\mn@eprint@dblp#1{\href {http://dblp.uni-trier.de/rec/bibtex/#1.xml}
  {dblp:#1}}
\def\mn@eprint@#1:#2:#3:#4\@nil{\def\@tempa {#1}\def\@tempb {#2}\def\@tempc
  {#3}\ifx \@tempc \@empty \let \@tempc \@tempb \let \@tempb \@tempa \fi \ifx
  \@tempb \@empty \def\@tempb {arXiv}\fi \@ifundefined
  {mn@eprint@\@tempb}{\@tempb:\@tempc}{\expandafter \expandafter \csname
  mn@eprint@\@tempb\endcsname \expandafter{\@tempc}}}

\bibitem[\protect\citeauthoryear{Achterberg, Gallant, Kirk  \&
  Guthmann}{Achterberg et~al.}{2001}]{2001MNRAS.328..393A}
Achterberg A.,  Gallant Y.~A.,  Kirk J.~G.,   Guthmann A.~W.,  2001, Monthly
  Notices of the Royal Astronomical Society, 328, 393

\bibitem[\protect\citeauthoryear{Aihara et~al.,}{Aihara
  et~al.}{2011}]{2011ApJS..193...29A}
Aihara H.,  et~al., 2011, The Astrophysical Journal Supplement Series, 193, 29

\bibitem[\protect\citeauthoryear{Barthelmy et~al.,}{Barthelmy
  et~al.}{2005}]{2005SSRv..120..143B}
Barthelmy S.~D.,  et~al., 2005, Space Science Reviews, 120, 143

\bibitem[\protect\citeauthoryear{Bertin \& Arnouts}{Bertin \&
  Arnouts}{1996}]{1996A&AS..117..393B}
Bertin E.,  Arnouts S.,  1996, Astronomy and Astrophysics Supplement, 117, 393

\bibitem[\protect\citeauthoryear{Bloom, Kulkarni  \& Djorgovski}{Bloom
  et~al.}{2002}]{2002AJ....123.1111B}
Bloom J.~S.,  Kulkarni S.~R.,   Djorgovski S.~G.,  2002, The Astronomical
  Journal, 123, 1111

\bibitem[\protect\citeauthoryear{Breeveld, Landsman, Holland, Roming, Kuin  \&
  Page}{Breeveld et~al.}{2011}]{2011AIPC.1358..373B}
Breeveld A.~A.,  Landsman W.,  Holland S.~T.,  Roming P.,  Kuin N. P.~M.,
  Page M.~J.,  2011, in GAMMA RAY BURSTS 2010. AIP Conference Proceedings. AIP,
  pp 373--376

\bibitem[\protect\citeauthoryear{Burrows et~al.,}{Burrows
  et~al.}{2005}]{2005SSRv..120..165B}
Burrows D.~N.,  et~al., 2005, Space Science Reviews, 120, 165

\bibitem[\protect\citeauthoryear{Cano et~al.,}{Cano
  et~al.}{2011}]{2011MNRAS.413..669C}
Cano Z.,  et~al., 2011, Monthly Notices of the Royal Astronomical Society, 413,
  669

\bibitem[\protect\citeauthoryear{Castro-Tirado et~al.,}{Castro-Tirado
  et~al.}{2001}]{2001A&A...370..398C}
Castro-Tirado A.~J.,  et~al., 2001, Astronomy and Astrophysics, 370, 398

\bibitem[\protect\citeauthoryear{Chandra \& Frail}{Chandra \&
  Frail}{2012}]{2012ApJ...746..156C}
Chandra P.,  Frail D.~A.,  2012, The Astrophysical Journal, 746, 156

\bibitem[\protect\citeauthoryear{Christensen, Fynbo, Prochaska, Th{\"o}ne, de
  Ugarte~Postigo  \& Jakobsson}{Christensen et~al.}{2011}]{2011ApJ...727...73C}
Christensen L.,  Fynbo J. P.~U.,  Prochaska J.~X.,  Th{\"o}ne C.~C.,  de
  Ugarte~Postigo A.,   Jakobsson P.,  2011, The Astrophysical Journal, 727, 73

\bibitem[\protect\citeauthoryear{Crowther}{Crowther}{2007}]{2007ARA&A..45..177C}
Crowther P.~A.,  2007, Annual Review of Astronomy and Astrophysics, 45, 177

\bibitem[\protect\citeauthoryear{Cucchiara et~al.,}{Cucchiara
  et~al.}{2011}]{2011ApJ...736....7C}
Cucchiara A.,  et~al., 2011, The Astrophysical Journal, 736, 7

\bibitem[\protect\citeauthoryear{D'Elia et~al.,}{D'Elia
  et~al.}{2014}]{2014A&A...564A..38D}
D'Elia V.,  et~al., 2014, Astronomy and Astrophysics, 564, A38

\bibitem[\protect\citeauthoryear{Evans et~al.,}{Evans
  et~al.}{2007}]{2007A&A...469..379E}
Evans P.~A.,  et~al., 2007, Astronomy and Astrophysics, 469, 379

\bibitem[\protect\citeauthoryear{Evans et~al.,}{Evans
  et~al.}{2009}]{2009MNRAS.397.1177E}
Evans P.~A.,  et~al., 2009, Monthly Notices of the Royal Astronomical Society,
  397, 1177

\bibitem[\protect\citeauthoryear{Evans et~al.,}{Evans
  et~al.}{2010}]{2010A&A...519A.102E}
Evans P.~A.,  et~al., 2010, Astronomy and Astrophysics, 519, A102

\bibitem[\protect\citeauthoryear{Evans, Goad, Osborne  \& Beardmore}{Evans
  et~al.}{2011}]{2011GCN..12161...1E}
Evans P.~A.,  Goad M.~R.,  Osborne J.~P.,   Beardmore A.~P.,  2011, GRB
  Coordinates Network, 1216

\bibitem[\protect\citeauthoryear{Feroz, Hobson  \& Bridges}{Feroz
  et~al.}{2009}]{2009MNRAS.398.1601F}
Feroz F.,  Hobson M.~P.,   Bridges M.,  2009, Monthly Notices of the Royal
  Astronomical Society, 398, 1601

\bibitem[\protect\citeauthoryear{Filgas et~al.,}{Filgas
  et~al.}{2011}]{2011A&A...526A.113F}
Filgas R.,  et~al., 2011, Astronomy and Astrophysics, 526, A113

\bibitem[\protect\citeauthoryear{Frail, Kulkarni, Nicastro, Feroci  \&
  Taylor}{Frail et~al.}{1997}]{1997Natur.389..261F}
Frail D.~A.,  Kulkarni S.~R.,  Nicastro L.,  Feroci M.,   Taylor G.~B.,  1997,
  Nature, 389, 261

\bibitem[\protect\citeauthoryear{Fynbo et~al.,}{Fynbo
  et~al.}{2009}]{2009ApJS..185..526F}
Fynbo J. P.~U.,  et~al., 2009, The Astrophysical Journal Supplement Series,
  185, 526

\bibitem[\protect\citeauthoryear{Galama et~al.,}{Galama
  et~al.}{1998}]{1998Natur.395..670G}
Galama T.~J.,  et~al., 1998, Nature, 395, 670

\bibitem[\protect\citeauthoryear{Gehrels et~al.,}{Gehrels
  et~al.}{2004}]{2004ApJ...611.1005G}
Gehrels N.,  et~al., 2004, The Astrophysical Journal, 611, 1005

\bibitem[\protect\citeauthoryear{Goldoni, Royer, Fran{\c c}ois, Horrobin,
  Blanc, Vernet, Modigliani  \& Larsen}{Goldoni
  et~al.}{2006}]{2006SPIE.6269E..2KG}
Goldoni P.,  Royer F.,  Fran{\c c}ois P.,  Horrobin M.,  Blanc G.,  Vernet J.,
  Modigliani A.,   Larsen J.,  2006, Ground-based and Airborne Instrumentation
  for Astronomy. Edited by McLean, 6269, 62692K

\bibitem[\protect\citeauthoryear{Golenetskii et~al.,}{Golenetskii
  et~al.}{2011}]{2011GCN..12166...1G}
Golenetskii S.,  et~al., 2011, GRB Coordinates Network, 1216, 1

\bibitem[\protect\citeauthoryear{Greiner et~al.,}{Greiner
  et~al.}{2008}]{2008PASP..120..405G}
Greiner J.,  et~al., 2008, The Publications of the Astronomical Society of the
  Pacific, 120, 405

\bibitem[\protect\citeauthoryear{Guidorzi et~al.,}{Guidorzi
  et~al.}{2014}]{2014MNRAS.438..752G}
Guidorzi C.,  et~al., 2014, Monthly Notices of the Royal Astronomical Society,
  438, 752

\bibitem[\protect\citeauthoryear{Hancock, Murphy  \& Schmidt}{Hancock
  et~al.}{2011}]{2011GCN..12171...1H}
Hancock P.~J.,  Murphy T.,   Schmidt B.~P.,  2011, GRB Coordinates Network,
  1217

\bibitem[\protect\citeauthoryear{Hartoog et~al.,}{Hartoog
  et~al.}{2015}]{2015A&A...580A.139H}
Hartoog O.~E.,  et~al., 2015, Astronomy and Astrophysics, 580, A139

\bibitem[\protect\citeauthoryear{Hjorth et~al.,}{Hjorth
  et~al.}{2003}]{2003Natur.423..847H}
Hjorth J.,  et~al., 2003, Nature, 423, 847

\bibitem[\protect\citeauthoryear{Horne}{Horne}{1986}]{1986PASP...98..609H}
Horne K.,  1986, The Publications of the Astronomical Society of the Pacific,
  98, 609

\bibitem[\protect\citeauthoryear{Jel{\'\i}nek et~al.,}{Jel{\'\i}nek
  et~al.}{2006}]{2006A&A...454L.119J}
Jel{\'\i}nek M.,  et~al., 2006, Astronomy and Astrophysics, 454, L119

\bibitem[\protect\citeauthoryear{J{\'o}hannesson, Bj{\"o}rnsson  \&
  Gudmundsson}{J{\'o}hannesson et~al.}{2006}]{2006ApJ...647.1238J}
J{\'o}hannesson G.,  Bj{\"o}rnsson G.,   Gudmundsson E.~H.,  2006, The
  Astrophysical Journal, 647, 1238

\bibitem[\protect\citeauthoryear{Johnston \& de Vegt}{Johnston \&
  de~Vegt}{1986}]{1986HiA.....7..103J}
Johnston K.~J.,  de Vegt C.,  1986, in IN: Highlights of astronomy. Volume 7 -
  Proceedings of the Nineteenth IAU General Assembly. pp 103--108

\bibitem[\protect\citeauthoryear{Kann, Klose  \& Zeh}{Kann
  et~al.}{2006}]{2006ApJ...641..993K}
Kann D.~A.,  Klose S.,   Zeh A.,  2006, The Astrophysical Journal, 641, 993

\bibitem[\protect\citeauthoryear{Kann et~al.,}{Kann
  et~al.}{2010}]{2010ApJ...720.1513K}
Kann D.~A.,  et~al., 2010, The Astrophysical Journal, 720, 1513

\bibitem[\protect\citeauthoryear{Kann et~al.,}{Kann
  et~al.}{2011}]{2011ApJ...734...96K}
Kann D.~A.,  et~al., 2011, The Astrophysical Journal, 734, 96

\bibitem[\protect\citeauthoryear{Kelson}{Kelson}{2003}]{2003PASP..115..688K}
Kelson D.~D.,  2003, The Publications of the Astronomical Society of the
  Pacific, 115, 688

\bibitem[\protect\citeauthoryear{Kistler, Y{\"u}ksel, Beacom, Hopkins  \&
  Wyithe}{Kistler et~al.}{2009}]{2009ApJ...705L.104K}
Kistler M.~D.,  Y{\"u}ksel H.,  Beacom J.~F.,  Hopkins A.~M.,   Wyithe J.
  S.~B.,  2009, The Astrophysical Journal, 705, L104

\bibitem[\protect\citeauthoryear{Klebesadel, Strong  \& Olson}{Klebesadel
  et~al.}{1973}]{1973ApJ...182L..85K}
Klebesadel R.~W.,  Strong I.~B.,   Olson R.~A.,  1973, The Astrophysical
  Journal, 182, L85

\bibitem[\protect\citeauthoryear{Kopa{\v c} et~al.,}{Kopa{\v c}
  et~al.}{2015}]{2015ApJ...813....1K}
Kopa{\v c} D.,  et~al., 2015, The Astrophysical Journal, 813, 1

\bibitem[\protect\citeauthoryear{Kouveliotou, Meegan, Fishman, Bhat, Briggs,
  Koshut, Paciesas  \& Pendleton}{Kouveliotou
  et~al.}{1993}]{1993ApJ...413L.101K}
Kouveliotou C.,  Meegan C.~A.,  Fishman G.~J.,  Bhat N.~P.,  Briggs M.~S.,
  Koshut T.~M.,  Paciesas W.~S.,   Pendleton G.~N.,  1993, The Astrophysical
  Journal, 413, L101

\bibitem[\protect\citeauthoryear{Kov{\'a}cs}{Kov{\'a}cs}{2008}]{2008SPIE.7020E..1SK}
Kov{\'a}cs A.,  2008, Millimeter and Submillimeter Detectors and
  Instrumentation for Astronomy IV. Edited by Duncan, 7020, 70201S

\bibitem[\protect\citeauthoryear{Kr{\"u}hler et~al.,}{Kr{\"u}hler
  et~al.}{2008}]{2008ApJ...685..376K}
Kr{\"u}hler T.,  et~al., 2008, The Astrophysical Journal, 685, 376

\bibitem[\protect\citeauthoryear{Kr{\"u}hler et~al.,}{Kr{\"u}hler
  et~al.}{2015}]{2015A&A...581A.125K}
Kr{\"u}hler T.,  et~al., 2015, Astronomy and Astrophysics, 581, A125

\bibitem[\protect\citeauthoryear{Laskar et~al.,}{Laskar
  et~al.}{2013}]{2013ApJ...776..119L}
Laskar T.,  et~al., 2013, The Astrophysical Journal, 776, 119

\bibitem[\protect\citeauthoryear{M{\'e}sz{\'a}ros \& Rees}{M{\'e}sz{\'a}ros \&
  Rees}{1993}]{1993ApJ...405..278M}
M{\'e}sz{\'a}ros P.,  Rees M.~J.,  1993, The Astrophysical Journal, 405, 278

\bibitem[\protect\citeauthoryear{Modigliani et~al.,}{Modigliani
  et~al.}{2010}]{2010SPIE.7737E..28M}
Modigliani A.,  et~al., 2010, in Silva D.~R.,  Peck A.~B.,   Soifer B.~T.,
  eds, Proceedings of the SPIE. SPIE, pp 773728--773728--12

\bibitem[\protect\citeauthoryear{Mundell et~al.,}{Mundell
  et~al.}{2013}]{2013Natur.504..119M}
Mundell C.~G.,  et~al., 2013, Nature, 504, 119

\bibitem[\protect\citeauthoryear{Nelson}{Nelson}{2011}]{2011GCN..12174...1N}
Nelson P.,  2011, GRB Coordinates Network, 1217

\bibitem[\protect\citeauthoryear{Oates et~al.,}{Oates
  et~al.}{2009}]{2009MNRAS.395..490O}
Oates S.~R.,  et~al., 2009, Monthly Notices of the Royal Astronomical Society,
  395, 490

\bibitem[\protect\citeauthoryear{Panaitescu \& Kumar}{Panaitescu \&
  Kumar}{2001}]{2001ApJ...554..667P}
Panaitescu A.,  Kumar P.,  2001, The Astrophysical Journal, 554, 667

\bibitem[\protect\citeauthoryear{Panaitescu \& Kumar}{Panaitescu \&
  Kumar}{2002}]{2002ApJ...571..779P}
Panaitescu A.,  Kumar P.,  2002, The Astrophysical Journal, 571, 779

\bibitem[\protect\citeauthoryear{Perley et~al.,}{Perley
  et~al.}{2014}]{2014ApJ...781...37P}
Perley D.~A.,  et~al., 2014, The Astrophysical Journal, 781, 37

\bibitem[\protect\citeauthoryear{Perley et~al.,}{Perley
  et~al.}{2016a}]{2016ApJ...817....7P}
Perley D.~A.,  et~al., 2016a, The Astrophysical Journal, 817, 7

\bibitem[\protect\citeauthoryear{Perley et~al.,}{Perley
  et~al.}{2016b}]{2016ApJ...817....8P}
Perley D.~A.,  et~al., 2016b, The Astrophysical Journal, 817, 8

\bibitem[\protect\citeauthoryear{Piranomonte, Vergani, Malesani, Fynbo,
  Wiersema  \& Kaper}{Piranomonte et~al.}{2011}]{2011GCN..12164...1P}
Piranomonte S.,  Vergani S.~D.,  Malesani D.,  Fynbo J. P.~U.,  Wiersema K.,
  Kaper L.,  2011, GRB Coordinates Network, 1216

\bibitem[\protect\citeauthoryear{Planck et~al.,}{Planck
  et~al.}{2014}]{2014A&A...571A..16P}
Planck C.,  et~al., 2014, Astronomy and Astrophysics, 571, A16

\bibitem[\protect\citeauthoryear{Racusin et~al.,}{Racusin
  et~al.}{2008}]{2008Natur.455..183R}
Racusin J.~L.,  et~al., 2008, Nature, 455, 183

\bibitem[\protect\citeauthoryear{Resmi et~al.,}{Resmi
  et~al.}{2012}]{2012MNRAS.427..288R}
Resmi L.,  et~al., 2012, Monthly Notices of the Royal Astronomical Society,
  427, 288

\bibitem[\protect\citeauthoryear{Rhoads}{Rhoads}{1997}]{1997ApJ...487L...1R}
Rhoads J.~E.,  1997, The Astrophysical Journal, 487, L1

\bibitem[\protect\citeauthoryear{Robertson \& Ellis}{Robertson \&
  Ellis}{2012}]{2012ApJ...744...95R}
Robertson B.~E.,  Ellis R.~S.,  2012, The Astrophysical Journal, 744, 95

\bibitem[\protect\citeauthoryear{Roming et~al.,}{Roming
  et~al.}{2005}]{2005SSRv..120...95R}
Roming P. W.~A.,  et~al., 2005, Space Science Reviews, 120, 95

\bibitem[\protect\citeauthoryear{Ryan, van Eerten, MacFadyen  \& Zhang}{Ryan
  et~al.}{2015}]{2015ApJ...799....3R}
Ryan G.,  van Eerten H.,  MacFadyen A.,   Zhang B.-B.,  2015, The Astrophysical
  Journal, 799, 3

\bibitem[\protect\citeauthoryear{Salvaterra et~al.,}{Salvaterra
  et~al.}{2009}]{2009Natur.461.1258S}
Salvaterra R.,  et~al., 2009, Nature, 461, 1258

\bibitem[\protect\citeauthoryear{Sari, Piran  \& Narayan}{Sari
  et~al.}{1998}]{1998ApJ...497L..17S}
Sari R.,  Piran T.,   Narayan R.,  1998, The Astrophysical Journal, 497, L17

\bibitem[\protect\citeauthoryear{Sari, Piran  \& Halpern}{Sari
  et~al.}{1999}]{1999ApJ...519L..17S}
Sari R.,  Piran T.,   Halpern J.~P.,  1999, The Astrophysical Journal, 519, L17

\bibitem[\protect\citeauthoryear{Sault, Teuben  \& Wright}{Sault
  et~al.}{1995}]{1995ASPC...77..433S}
Sault R.~J.,  Teuben P.~J.,   Wright M. C.~H.,  1995, in Astronomical Data
  Analysis Software and Systems IV. p.~433

\bibitem[\protect\citeauthoryear{Schlafly \& Finkbeiner}{Schlafly \&
  Finkbeiner}{2011}]{2011ApJ...737..103S}
Schlafly E.~F.,  Finkbeiner D.~P.,  2011, The Astrophysical Journal, 737, 103

\bibitem[\protect\citeauthoryear{Schlegel, Finkbeiner  \& Davis}{Schlegel
  et~al.}{1998}]{1998ApJ...500..525S}
Schlegel D.~J.,  Finkbeiner D.~P.,   Davis M.,  1998, The Astrophysical
  Journal, 500, 525

\bibitem[\protect\citeauthoryear{Schulze et~al.,}{Schulze
  et~al.}{2015}]{2015ApJ...808...73S}
Schulze S.,  et~al., 2015, The Astrophysical Journal, 808, 73

\bibitem[\protect\citeauthoryear{Shen, Kumar  \& Robinson}{Shen
  et~al.}{2006}]{2006MNRAS.371.1441S}
Shen R.,  Kumar P.,   Robinson E.~L.,  2006, Monthly Notices of the Royal
  Astronomical Society, 371, 1441

\bibitem[\protect\citeauthoryear{Siringo et~al.,}{Siringo
  et~al.}{2009}]{2009A&A...497..945S}
Siringo G.,  et~al., 2009, Astronomy and Astrophysics, 497, 945

\bibitem[\protect\citeauthoryear{Skrutskie et~al.,}{Skrutskie
  et~al.}{2006}]{2006AJ....131.1163S}
Skrutskie M.~F.,  et~al., 2006, The Astronomical Journal, 131, 1163

\bibitem[\protect\citeauthoryear{Sonbas et~al.,}{Sonbas
  et~al.}{2011}]{2011GCN..12158...1S}
Sonbas E.,  et~al., 2011, GRB Coordinates Network, 1215

\bibitem[\protect\citeauthoryear{Sparre et~al.,}{Sparre
  et~al.}{2014}]{2014ApJ...785..150S}
Sparre M.,  et~al., 2014, The Astrophysical Journal, 785, 150

\bibitem[\protect\citeauthoryear{Starling, Wijers, Hughes, Tanvir, Vreeswijk,
  Rol  \& Salamanca}{Starling et~al.}{2005}]{2005MNRAS.360..305S}
Starling R. L.~C.,  Wijers R. A. M.~J.,  Hughes M.~A.,  Tanvir N.~R.,
  Vreeswijk P.~M.,  Rol E.,   Salamanca I.,  2005, Monthly Notices of the Royal
  Astronomical Society, 360, 305

\bibitem[\protect\citeauthoryear{Tanvir et~al.,}{Tanvir
  et~al.}{2009}]{2009Natur.461.1254T}
Tanvir N.~R.,  et~al., 2009, Nature, 461, 1254

\bibitem[\protect\citeauthoryear{Th{\"o}ne et~al.,}{Th{\"o}ne
  et~al.}{2010}]{2010A&A...523A..70T}
Th{\"o}ne C.~C.,  et~al., 2010, Astronomy and Astrophysics, 523, A70

\bibitem[\protect\citeauthoryear{Tinney et~al.,}{Tinney
  et~al.}{1998}]{1998IAUC.6896....3W}
Tinney C.,  et~al., 1998, Circulars of the International Astronomical Union,
  6896, 1

\bibitem[\protect\citeauthoryear{Tody}{Tody}{1993}]{1993ASPC...52..173T}
Tody D.,  1993, Astronomical Data Analysis Software and Systems II, 52, 173

\bibitem[\protect\citeauthoryear{Updike, Schady, Greiner, Kr{\"u}hler, Kann,
  Klose  \& Rossi}{Updike et~al.}{2011}]{2011GCN..12169...1U}
Updike A.~C.,  Schady P.,  Greiner J.,  Kr{\"u}hler T.,  Kann D.~A.,  Klose S.,
    Rossi A.,  2011, GRB Coordinates Network, 1216

\bibitem[\protect\citeauthoryear{Vergani et~al.,}{Vergani
  et~al.}{2015}]{2015A&A...581A.102V}
Vergani S.~D.,  et~al., 2015, Astronomy and Astrophysics, 581, A102

\bibitem[\protect\citeauthoryear{Vernet et~al.,}{Vernet
  et~al.}{2011}]{2011A&A...536A.105V}
Vernet J.,  et~al., 2011, Astronomy and Astrophysics, 536, A105

\bibitem[\protect\citeauthoryear{Walker}{Walker}{1998}]{1998MNRAS.294..307W}
Walker M.~A.,  1998, Monthly Notices of the Royal Astronomical Society, 294,
  307

\bibitem[\protect\citeauthoryear{Wijers \& Galama}{Wijers \&
  Galama}{1999}]{1999ApJ...523..177W}
Wijers R. A. M.~J.,  Galama T.~J.,  1999, The Astrophysical Journal, 523, 177

\bibitem[\protect\citeauthoryear{Wilson et~al.,}{Wilson
  et~al.}{2011}]{2011MNRAS.416..832W}
Wilson W.~E.,  et~al., 2011, Monthly Notices of the Royal Astronomical Society,
  416, 832

\bibitem[\protect\citeauthoryear{Woosley \& Bloom}{Woosley \&
  Bloom}{2006}]{2006ARA&A..44..507W}
Woosley S.~E.,  Bloom J.~S.,  2006, Annual Review of Astronomy and
  Astrophysics, 44, 507

\bibitem[\protect\citeauthoryear{Yolda{\c s}, Kr{\"u}hler, Greiner, Yolda{\c
  s}, Clemens, Szokoly, Primak  \& Klose}{Yolda{\c s}
  et~al.}{2008}]{2008AIPC.1000..227Y}
Yolda{\c s} A.~K.,  Kr{\"u}hler T.,  Greiner J.,  Yolda{\c s} A.,  Clemens C.,
  Szokoly G.,  Primak N.,   Klose S.,  2008, in GAMMA-RAY BURSTS 2007:
  Proceedings of the Santa Fe Conference. AIP Conference Proceedings. AIP, pp
  227--231

\bibitem[\protect\citeauthoryear{Zhang \& Yan}{Zhang \&
  Yan}{2011}]{2011ApJ...726...90Z}
Zhang B.,  Yan H.,  2011, The Astrophysical Journal, 726, 90

\bibitem[\protect\citeauthoryear{Zhang, Zhang, Murase, Connaughton  \&
  Briggs}{Zhang et~al.}{2014}]{2014ApJ...787...66Z}
Zhang B.-B.,  Zhang B.,  Murase K.,  Connaughton V.,   Briggs M.~S.,  2014, The
  Astrophysical Journal, 787, 66

\bibitem[\protect\citeauthoryear{Zhang, van Eerten, Burrows, Ryan, Evans,
  Racusin, Troja  \& MacFadyen}{Zhang et~al.}{2015}]{2015ApJ...806...15Z}
Zhang B.-B.,  van Eerten H.,  Burrows D.~N.,  Ryan G.~S.,  Evans P.~A.,
  Racusin J.~L.,  Troja E.,   MacFadyen A.,  2015, The Astrophysical Journal,
  806, 15

\bibitem[\protect\citeauthoryear{Zheng et~al.,}{Zheng
  et~al.}{2012}]{2012ApJ...751...90Z}
Zheng W.,  et~al., 2012, The Astrophysical Journal, 751, 90

\bibitem[\protect\citeauthoryear{de Ugarte~Postigo et~al.,}{de~Ugarte~Postigo
  et~al.}{2006}]{2006ApJ...648L..83D}
de Ugarte~Postigo A.,  et~al., 2006, The Astrophysical Journal, 648, L83

\bibitem[\protect\citeauthoryear{de Ugarte~Postigo et~al.,}{de~Ugarte~Postigo
  et~al.}{2007}]{2007A&A...462L..57D}
de Ugarte~Postigo A.,  et~al., 2007, Astronomy and Astrophysics, 462, L57

\bibitem[\protect\citeauthoryear{de Ugarte~Postigo et~al.,}{de~Ugarte~Postigo
  et~al.}{2011}]{2011GCN..12168...1D}
de Ugarte~Postigo A.,  et~al., 2011, GRB Coordinates Network, 1216, 1

\bibitem[\protect\citeauthoryear{de Ugarte~Postigo et~al.,}{de~Ugarte~Postigo
  et~al.}{2012a}]{2012A&A...538A..44D}
de Ugarte~Postigo A.,  et~al., 2012a, Astronomy and Astrophysics, 538, A44

\bibitem[\protect\citeauthoryear{de Ugarte~Postigo et~al.,}{de~Ugarte~Postigo
  et~al.}{2012b}]{2012A&A...548A..11D}
de Ugarte~Postigo A.,  et~al., 2012b, Astronomy and Astrophysics, 548, A11

\bibitem[\protect\citeauthoryear{van Dokkum}{van
  Dokkum}{2001}]{2001PASP..113.1420V}
van Dokkum P.~G.,  2001, The Publications of the Astronomical Society of the
  Pacific, 113, 1420

\bibitem[\protect\citeauthoryear{van Eerten, van~der Horst  \& MacFadyen}{van
  Eerten et~al.}{2012}]{2012ApJ...749...44V}
van Eerten H.,  van~der Horst A.,   MacFadyen A.,  2012, The Astrophysical
  Journal, 749, 44

\bibitem[\protect\citeauthoryear{van~der Horst et~al.,}{van~der Horst
  et~al.}{2014}]{2014MNRAS.444.3151V}
van~der Horst A.~J.,  et~al., 2014, Monthly Notices of the Royal Astronomical
  Society, 444, 3151

\makeatother
\end{thebibliography}

\newpage
\noindent
$^{1}$Instituto de Astrof\'{\i}sica de Andaluc\'{\i}a (IAA-CSIC), Glorieta de la Astronom\'{\i}a s/n, E-18008, Granada, Spain.\\
$^{2}$Unidad Asociada Grupo Ciencias Planetarias (UPV/EHU, IAA-CSIC), Departamento de F\'{\i}sica Aplicada I, E.T.S. Ingenier\'{\i}a, Universidad del Pa\'{\i}s Vasco (UPV/EHU), Alameda de Urquijo s/n, E-48013 Bilbao, Spain.\\
$^{3}$Ikerbasque, Basque Foundation for Science, Alameda de Urquijo 36-5, E-48008 Bilbao, Spain.\\
$^{4}$Sydney Institute for Astronomy, School of Physics, The University of Sydney, NSW 2006, Australia.\\
$^{5}$ARC Centre of Excellence for All-sky Astrophysics (CAASTRO)  Australia.\\
$^{6}$Science Institute, University of Iceland, IS-107 Reykjavik, Iceland.\\
$^{7}$Dark Cosmology Centre, Niels Bohr Institute, Juliane Maries Vej 30, 2100 Copenhagen \O{}, Denmark.\\
$^{8}$Th\"{u}ringer Landessternwarte Tautenburg, Sternwarte 5, 07778 Tautenburg, Germany.\\
$^{9}$Max-Planck-Institut f\"{u}r extraterrestrische Physik, Giessenbachstra\ss e 1, 85748 Garching, Germany.\\
$^{10}$European Southern Observatory, Alonso de C\'{o}rdova 3107, Vitacura, Santiago, Chile.\\
$^{11}$Mullard Space Science Laboratory, University College London, Holmbury St. Mary, Dorking, Surrey RH5 6NT, UK.\\
$^{12}$INAF - Osservatorio Astronomico di Trieste, via G. B. Tiepolo 11, 34131 Trieste, Italy.\\
$^{13}$Joint ALMA Observatory, Alonso de C\'{o}rdova 3107, Vitacura, Santiago, Chile.\\
$^{14}$INAF-Osservatorio Astronomico di Roma, via Frascati 33, I-00040 Monte Porzio Catone (RM), Italy.\\
$^{15}$ASI-Science Data Center, Via del Politecnico snc, I-00133 Rome, Italy.\\
$^{16}$APC, Univ. Paris Diderot, CNRS/IN2P3, CEA/Irfu, Obs. de Paris, Sorbonne Paris Cit, France.\\
$^{17}$Universe Cluster, Technische Universit\"{a}t M\"{u}nchen, Boltzmannstra\ss e 2, 85748 Garching, Germany.\\
$^{18}$Astronomical Institute of the Czech Academy of Sciences, Fri\v{c}ova 298, CZ-25165 Ond\v{r}ejov, Czech Republic.\\
$^{19}$Institute of Science and Technology in Space, SungKyunKwan University, Suwon 16419, Republic of Korea.\\
$^{20}$Harvard-Smithsonian Center for Astrophysics, 60 Garden St., Cambridge, MA 02138, USA.\\
$^{21}$European Space Astronomy Center, ISDEFE, ESA, Villafranca del Castillo, 50727, 28080, Madrid, Spain.\\
$^{22}$INAF-IASF Bologna, Area della Ricerca CNR, via Gobetti 101, I--40129 Bologna, Italy.\\
$^{23}$Astrophysics Data System, Harvard-Smithonian Center for Astrophysics, Garden St. 60, Cambridge, MA 02138, USA.\\
$^{24}$Department of Chemistry and Physics, Roger Williams University, One Old Ferry Road, Bristol, RI 02809, USA.\\
$^{25}$Department of Physics and Astronomy, University of Leicester, University Road, Leicester LE1 7RH, UK.\\
$^{26}$Scientist Support LLC, Madison, AL 35758, USA.


\appendix

\newpage

\section{Global view of all performed fits}
\label{app:view}

\begin{table*}
\caption{The Bayesian evidence and reduced chi-square for all ($\chi^2_{red}$), X-ray ($\chi^2_{red, x}$), optical ($\chi^2_{red, o}$), and radio ($\chi^2_{red, r}$) bands. A detailed table for each model is available in the online version.}
\label{tab:goodness}
\begin{tabular}{llllll}
\hline\hline
Fit Id & Evidence & $\chi^2_{red}$ & $\chi^2_{red, x}$ & $\chi^2_{red, o}$ & $\chi^2_{red, r}$ \\
\hline
CM/XOR/WC & $-1066.52 \pm 0.19$ & 6.52 & 2.45 & 16.49 & 27.98 \\
TS/XOR/WC & $-951.65 \pm 0.20$ & 6.24 & 2.0 & 16.12 & 30.48 \\
WM/XOR/WC & $-1181.09 \pm 0.19$ & 9.78 & 2.61 & 29.6 & 39.61 \\
CM/XUOR/WN & $-1677.46 \pm 0.20$ & 6.92 & 2.46 & 15.75 & 35.57 \\
TS/XUOR/WN & $-1503.57 \pm 0.20$ & 6.79 & 1.92 & 18.22 & 31.42 \\
WM/XUOR/WN & $-1753.24 \pm 0.20$ & 57.02 & 2.25 & 259.83 & 47.46 \\
CM/XOR/WN & $-1667.78 \pm 0.20$ & 6.79 & 2.45 & 15.86 & 35.35 \\
TS/XOR/WN & $-1491.05 \pm 0.20$ & 6.5 & 1.91 & 17.61 & 31.27 \\
WM/XOR/WN & $-1730.71 \pm 0.20$ & 54.46 & 2.21 & 261.71 & 47.0 \\
CM/XOR/VC & $-1039.59 \pm 0.20$ & 5.69 & 2.49 & 12.62 & 26.91 \\
TS/XOR/VC & $-988.59 \pm 0.20$ & 5.51 & 2.14 & 11.86 & 31.2 \\
WM/XOR/VC & $-1145.41 \pm 0.19$ & 7.45 & 2.73 & 18.03 & 37.6 \\
CM/XOR/VN & $-1015.83 \pm 0.21$ & 5.73 & 2.47 & 13.0 & 26.79 \\
TS/XOR/VN & $-995.70 \pm 0.20$ & 5.5 & 2.15 & 11.34 & 32.54 \\
WM/XOR/VN & $-1129.24 \pm 0.19$ & 7.3 & 2.76 & 17.48 & 36.38 \\
CM/O/VC & $-201.49 \pm 0.16$ & 175.87 & 8.82 & 37.34 & 2997.26 \\
TS/O/VC & $-206.60 \pm 0.18$ & 71.38 & 19.67 & 70.76 & 800.15 \\
WM/O/VC & $-237.56 \pm 0.15$ & 727.47 & 277.16 & 228.39 & 8762.93 \\
CM/R & $-267.53 \pm 0.16$ & 57.98 & 48.6 & 100.81 & 25.24 \\
TS/R & $-252.66 \pm 0.15$ & 195.24 & 135.6 & 457.94 & 23.98 \\
WM/R & $-249.27 \pm 0.18$ & 467.74 & 190.45 & 1597.79 & 22.54 \\
CM/XUO/VC & $-555.22 \pm 0.18$ & 164.1 & 1.88 & 13.01 & 3000.39 \\
TS/XUO/VC & $-466.51 \pm 0.21$ & 53.66 & 1.73 & 68.71 & 727.94 \\
WM/XUO/VC & $-733.27 \pm 0.18$ & 183.79 & 2.12 & 14.73 & 3359.54 \\
CM/XR/ & $-600.80 \pm 0.18$ & 8.4 & 2.38 & 25.9 & 25.76 \\
TS/XR/ & $-558.08 \pm 0.18$ & 8.82 & 1.87 & 29.13 & 28.49 \\
WM/XR/ & $-671.99 \pm 0.18$ & 10.69 & 2.29 & 35.22 & 34.55 \\
CM/UOR/VC & $-561.73 \pm 0.19$ & 20.83 & 18.66 & 26.5 & 30.35 \\
TS/UOR/VC & $-541.22 \pm 0.19$ & 47.79 & 41.7 & 77.02 & 25.68 \\
WM/UOR/VC & $-676.66 \pm 0.19$ & 23.52 & 25.91 & 11.13 & 35.58 \\
CM/XUOR/VCE & $-848.73 \pm 0.19$ & 9.82 & 3.66 & 21.3 & 28.05 \\
TS/XUOR/VCE & $-670.00 \pm 0.20$ & 13.16 & 2.24 & 39.29 & 29.35 \\
WM/XUOR/VCE & $-1044.04 \pm 0.19$ & 15.01 & 3.05 & 38.75 & 46.29 \\
CM/XUOR/VCL & $-547.08 \pm 0.18$ & 14.02 & 1.65 & 26.9 & 28.55 \\
TS/XUOR/VCL & $-494.77 \pm 0.19$ & 20.93 & 1.86 & 47.0 & 29.91 \\
WM/XUOR/VCL & $-672.83 \pm 0.18$ & 65.64 & 2.99 & 174.44 & 45.23 \\
\hline
\end{tabular}
\\[0.2cm]
\begin{scriptsize}

Fit Id = (A)/(B)/(C)(D)(E)\\

(A) = Model used (CM/TS/WM)\\

(B) = Wavelength range of the observations used for the model fitting:\\
\hspace{0.2cm} X = XRT 2 keV\\
\hspace{0.2cm} U = UVOT UVW2, UVM2, UVW1, and U\\
\hspace{0.2cm} O = Rest of the UVOT and GROND bands\\
\hspace{0.2cm} R = Radio and sub-mm bands\\

(C) = UVOT \textit{white} band shifted to UVOT (V) or independent (W)\\

(D) = Treatment of the Galactic reddening:\\
\hspace{0.2cm} C = Corrected\\
\hspace{0.2cm} N = Set to a nuisance parameter\\

(E) = Time interval used for the model fitting:\\
\hspace{0.2cm} E = From $t$ = 0.05 days\\
\hspace{0.2cm} L = From $t$ = 0.5 days\\

\end{scriptsize}
\end{table*}

The goodness of the complete grid of fits are summarised in Table \ref{tab:goodness}. This grid consists of different cuts in wavelength and time in order to get useful additional information that clarifies some details of the physical nature of the afterglow and systematic uncertainties.

To test the constraining power of each wavelength range, we split the data into three subsets: X-rays, [UV/]optical/nIR\footnote{We include in the UV UVOT filters uvw1, uvw2, uvm2, and u}, and submm/mm. These were then fitted individually and also in sets of two.

We found that the optical light curves were best fit with the TS and WM model, where the early steep decay and the bump at around 0.3 days is easily explained. The TS model was slightly better, mostly due to a better fit to the early \textit{white+v} band data. The TS model also does a better job of predicting the XRT and radio data while the WM model is orders of magnitude off. The CM model does not do as well with the optical data, a large energy injection in combination with a low value for $p$ does a reasonable job at explaining the late light curves, but the early \textit{white+v} band data is not explained. The CM model also fares better with predicting the XRT and radio/sub-mm data, although it is obviously not able to reproduce them completely.

The fit to the radio data is less discriminating, the WM models are better than both the TS and CM models, but only marginally. The parameters for the WM and TS models are very similar and the energy injection and wind termination shock both happen at late times to improve the fit to the late radio points. The CM model stands out from the group with the energy injection happening at early times and is therefore the worst offender at late times. The CM model, however, is best at predicting the optical and X-ray data and roughly goes through the late time optical/nIR curves and the XRT curves. The WM and TS model under-predict those same data, with the WM model being the worst offender. Early observations would helped in constraining better modeled light curves.

No attempt was made at fitting the XRT light curve only, but when we add it to the mix with either the optical/NIR or the radio/sub-mm data things change considerably. For the former set it is now the TS model that is best, trailed by the CM and then the WM model. None of the models now explain the bump in the optical light curves, but at least the TS model explains the wiggles in the XRT light curve. The early \textit{white+v} optical data are also not explained. In this case, the TS and WM models do a fairly good job of predicting the radio/sub-mm data, but the CM model is way off. For the latter set of XRT and radio/sub-mm we get a pretty consistent picture of the three models. The TS model is best, trailed by the CM model and finally the WM model like for the entire set. The resulting parameter distributions are actually fairly close to the results of the entire dataset, indicating that the additional information from the
optical data does not constrain the model much. All of the models actually predict the optical data reasonably well and the full fit gives only small visible changes. This means that the large spectral lever arm added to the very fine temporal sampling of the XRT light curves is most constraining for the model.

Our final combination is the UV/radio/sub-mm and optical/NIR data together. Here the CM model shows the best fit, which fairs similarly to the WM, and significantly better fit than TS model. None of them is able to explain the bump in optical, but the early optical and early radio/sub-mm data is well explained by the TS model. All of the models approximately predict the XRT light curve but with some offset in the temporal behaviour. There is therefore little additional constraining power in the spectral information from the XRT data, but mostly from the very detailed time behavior.

In conclusion, it seems that the fine sampling in the XRT light curve with the large spectral lever arm of the radio and sub-mm data is the most constraining data for the models. We also note that the inferred physical conditions can vary up to few orders of magnitude depending on the model and the wavelength ranges considered. Therefore, observational sampling is fundamental in order to discriminate different models and constrain its physical parameters.

To reduce the bias from the early UVOT \textit{white+v} band points that can be caused by a reverse shock \citep[RS; e.g.,][]{2006A&A...454L.119J}, we redo the analysis with all optical points before 0.005 days turned into upper limits. We note that a RS can also contribute at later times in sub-mm and mm bands, \citep[e.g.,][]{2013ApJ...776..119L,2014ApJ...781...37P}. However, it is not needed to include them for the qualitative overview we want to give. The parameters of the models are mostly unchanged with this exclusions of the data. A notable exception are the values of $\epsilon_e$, $\epsilon_i$, and $\epsilon_B$. $\epsilon_B$ is reduced significantly while both $\epsilon_e$ and $\epsilon_i$ increase. This affects the determined host extinction which is now determined to be twice as large. The models are still unable to reproduce the data and most of the comments still apply. The contribution of a reverse shock may help to explain the early evolution in the \textit{white} filter and the early sub-mm (and maybe mm) light curves, but will not help with the rest of the data. Thus the early \textit{white} band data is not the driving cause for the models not being able to reproduce the bump.

One possibility that has often been proposed to model complex GRB light curves is the double jet model \citep[e.g.][]{2005MNRAS.360..305S, 2008Natur.455..183R, 2011A&A...526A.113F, 2014MNRAS.444.3151V}, the early light curve being dominated by a fast moving narrow jet while a slow moving wide jet dominates at late time. This can be considered the simplest model for a two dimensional jet. To test if this is the case here, we fit the data after 0.5 days only, turning all other points into upper limits. The TS model is still the best model in this case and it is mostly able to explain the optical and X-ray bump at 0.3 days, but all the other considerations still apply and the radio/sub-mm data is still poorly modelled. We therefore conclude that a double jet model is not appropriate for this case. 

\bsp	
\label{lastpage}
\end{document}